%% file: kpwedge_arxiv.tex
\documentclass[aps,pre,superscriptaddress,amsmath,amssymb,amsfonts,twocolumn,showpacs,10pt]{revtex4-1}
\usepackage{amsmath,amssymb,amsfonts,graphicx,physics}
\usepackage{fourier,times}
\usepackage[colorlinks=true,citecolor=blue]{hyperref}
\newcommand\orcid[1]{\href{https://orcid.org/#1}{\includegraphics[width=10pt]{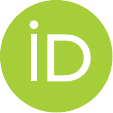}}}
\graphicspath{{figures/}}

\def\Real{\mathbb{R}}
\def\d{\mathrm{d}}
\def\F{\mathcal{F}}
\def\const{\mathrm{const}}
\def\be{\begin{equation}}
\def\ee{\end{equation}}
\def\bse{\begin{subequations}}
\def\ese{\end{subequations}}
\let\^=\hat
\def\cn{\mathop{\rm cn}\nolimits}
\def\min{\mathrm{min}}
\def\max{\mathrm{max}}
\def\crit{\mathrm{cr}}

\advance\textwidth 4mm
\advance\hoffset -2mm
\advance\textheight 4mm
\advance\voffset -2mm
\advance\parindent 0.4em
\advance\parskip 1pt

\def\thetitle{Mach reflection and expansion of two-dimensional dispersive shock waves}

\begin{document}
\title{\thetitle}

\author{Gino Biondini\orcid{0000-0003-3835-1343}}
\affiliation{Department of Mathematics, State University of New York, Buffalo, New York, United States of America}
\affiliation{Department of Physics, State University of New York, Buffalo, New York, United States of America}

\author{Alexander Bivolcic\orcid{0009-0001-7406-7692}}
\affiliation{Department of Mathematics, State University of New York, Buffalo, New York, United States of America}

\author{Mark A. Hoefer\orcid{0000-0001-5883-6562}}
\affiliation{Department of Applied Mathematics, University of Colorado, Boulder, Colorado, United States of America}
\date{\small\today}

\begin{abstract}
The oblique collisions and dynamical interference patterns of two-dimensional dispersive shock waves are studied numerically and analytically via the temporal dynamics induced by wedge-shaped initial conditions for the Kadomtsev-Petviashvili II equation.  
Various asymptotic wave patterns are identified, classified and characterized in terms of the incidence angle and the amplitude of the initial step, which can give rise to either subcritical or supercritical configurations,
including the generalization to dispersive shock waves of the Mach reflection and expansion of viscous shocks and line solitons. 
An eightfold amplification of the amplitude of an obliquely incident flow upon a wall at the critical angle is demonstrated.
\end{abstract}

\maketitle


\textbf{Introduction.}
The far-from-equilibrium, strongly nonlinear process of wavebreaking
is fundamental to the understanding of photonics \cite{wan_dispersive_2007,xu_dispersive_2017,bienaime_quantitative_2021,bendahmane_piston_2022}, quantum superfluids \cite{dutton_observation_2001,hoefer_dispersive_2006,joseph_observation_2011,mossman_dissipative_2018}, geophysics \cite{nash_river_2005,scotti_generation_2007,trillo_experimental_2016,li_seasonal_2018}, and many other wave systems \cite{mo_experimental_2013,maiden_observation_2016, janantha_observation_2017,li_observation_2021}.  
Whether referred to as an optical shock, a quantum shock, or an undular bore, wavebreaking in all of these scenarios results in a dispersive shock wave (DSW).  
DSWs consist of a multiscale, unsteady, coherent collection of nonlinear waves that are distinctly different and inherently more complex than their more familiar gas dynamic, viscous shock wave counterparts. 
While significant progress on experiments \cite{xu_dispersive_2017,bienaime_quantitative_2021,bendahmane_piston_2022,trillo_experimental_2016,mo_experimental_2013,maiden_observation_2016,janantha_observation_2017,li_observation_2021}, modeling \cite{trillo_observation_2016,chassagne_dispersive_2019,simmons_what_2020}, and analysis \cite{PHYSD333p11,miller_generation_2016} of one-dimensional (1D) DSWs has been achieved in recent years, the theory of multi-dimensional DSWs has been limited to symmetric \cite{ablowitz_dispersive_2016} or lower-dimensional/time-independent \cite{gurevich_supersonic_1995,gurevich_supersonic_1996,el_two-dimensional_2009,hoefer_dark_2012,hoefer_oblique_2017} reductions. 
Since most physical systems are intrinsically multi-dimensional, the study of DSWs in these systems is therefore warranted.

The simplest generation mechanism for shock waves is a Riemann problem, which consists in step initial conditions (ICs) for the hydrodynamic variables \cite{Lax1973,CF1976,CH1962}.  In two spatial dimensions (2D), the natural generalization of the Riemann problem is to consider ICs that are piecewise constant in different sectors of the plane \cite{zhang_conjecture_1990,kurganov_solution_2002}.  Such a scenario forms the foundation for the theoretical study of regular and Mach reflection in which an obliquely incident shock encounters a wall \cite{JFM79p171,brio_mach_1992}.  
In fact, the study of nonlinear wave reflection goes back to Russell's ``report on waves'' \cite{Russell1844}, which described experiments on the oblique reflection of a water wave soliton and observed what he termed ``lateral accumulation''.  Now known as Mach reflection, this general nonlinear process leads to the generation of a triple point where a stem wave connects to the incident and reflected waves.  While there has been a great deal of study on the Mach reflection of viscous shocks and oblique solitons, there has been no study of this fundamental, multi-dimensional nonlinear wave process for DSWs.

In this Letter we consider sectorial Riemann problems for the Kadomtsev-Petviashvili (KP) equation leading to the generation of two obliquely colliding DSWs, a problem that is related to the oblique interaction of a DSW with a wall.  
Numerical simulations for both compressive (acute) and expansive (obtuse) angles reveal a bifurcation between two distinctive behaviors, which we identify as the bifurcation from Mach reflection or expansion (subcritical) to regular reflection or expansion (supercritical), respectively, of a DSW obliquely incident upon a wall.  
We analytically identify salient properties of the unsteady, two-dimensional DSW patterns and their bifurcations including the critical angle, the leading-edge structure, the growth of the DSW Mach stem,
and up to an eightfold amplification of the amplitude of the oblique, incident flow.

\medskip
\textbf{Kadomtsev-Petviashvili equation.}
The KP equation~\cite{SovPhysDoklady15p539}\break
is a prototypical universal model for weakly two-dimensional, weakly nonlinear waves. 
It arises in a number of disparate physical settings such as shallow water \cite{AblowitzSegur1981,Kodama2018}, 
internal waves \cite{ablowitz_evolution_1979},
plasmas \cite{InfeldRowlands}, 
magnetic materials \cite{JETP62p146,JPA35p10149}, 
cosmology \cite{AA194p3,CQG19p2793,OC181p345,PLA158p107},
and Bose-Einstein condensates \cite{CPL19p17,PRE65p174518,PRA69p53601}.
In rescaled and dimensionless variables, it can be written as
\vspace*{-0.8ex}
\be
(u_t + u u_x + u_{xxx} )_ x + \,u_{yy} = 0\,.
\label{e:KP}
\ee
The meaning of the variable $u$ depends on the physical context considered, 
while $x$ represents the main direction of propagation, $y$ a transverse spatial variable
and $t$ is time.
In shallow water, $u$ quantifies the deviation of the water surface from its equilibrium value.  
The KP equation is also the prototypical infinite-dimensional completely integrable system in two spatial dimensions
and, as such, it enjoys a deep mathematical structure, 
including a Lax pair, inverse scattering transform
\cite{NMPZ1984,AblowitzClarkson1991,Konopelchenko1993}, 
as well as a large family of line-soliton solutions displaying phenomena of 
soliton resonance and web structure \cite{JPA36p10519,JMP47p033514,PRL99p064103,Kodama2017}.
Such integrable structure will play no role in the present analysis, however, so that the results are indicative of a broad class of physical systems where integrability may not apply.
Note that the KP equation comes in two variants, commonly referred to as KPI and KPII,
and Eq.~\eqref{e:KP} is the KPII variant.
(For the KPI equation, the term $u_{yy}$ would have a negative coefficient.)
In the water waves context, 
the KPII and KPI equation describes the case with weak and strong surface tension, respectively \cite{AblowitzSegur1981}.
The two variants of the equation differ both in their mathematical properties and in the physical behavior of solutions.
Here we only discuss the KPII equation, since only in this case the line soliton solutions are stable \cite{SovPhysDoklady15p539}.
Importantly, even though Eq.~\eqref{e:KP} has been the subject of a large number of studies over the last fifty years,
the temporal dynamics of solutions arising from generic ICs
is still not well understood.

For solutions that are independent of $y$, 
Eq.~\eqref{e:KP} reduces to the Korteweg-de\,Vries (KdV) equation
$u_t + u u_x + u_{xxx} = 0$.  
In turn, the KdV equation is a dispersive regularization of the Hopf equation
$u_t + u u_x  = 0$.
It is well known that increasing ICs for the Hopf equation give rise to rarefaction waves,  while decreasing ICs give rise to shocks \cite{Whitham1974}.
The dispersive regularization of the Hopf equation via the KdV equation for decreasing ICs gives rise instead to 
a DSW
\cite{PHYSD333p11}.

\medskip
\textbf{Line solitons, cnoidal waves and slanted DSWs.}
The KP equation~\eqref{e:KP} admits a three-parameter family of one-soliton solutions representing a solitary wave with amplitude~$a$ and slope~$q$ on a background $\bar u$, given by
\vspace*{-0.6ex}
\be
u(x,y,t) = \bar u + a\sech^2\bigg(\sqrt{\frac{a}{12}}(x+qy-ct-x_o)\bigg),
\ee
where $q=\tan\phi$ quantifies the transverse inclination angle $\phi$ of the soliton and 
$c= C(a,q,\bar u) = \bar u +a/3 + q^2$ is the soliton propagation velocity along the 
$x$ direction \cite{NLTY2021v34p3583}.
The solution is localized along $x+qy-ct = x_o$, which describes a (moving) line in the
$xy$-plane, and is therefore referred to as a \textit{line soliton}.
The KP equation also admits a ``slanted'' version of the DSWs of the KdV equation.
These are asymptotically described as a slow spatio-temporal modulation of the periodic traveling wave solutions
of the KP equation, which are given by 
\vspace*{-0.2ex}
\be
u(x,y,t) = r_1 - r_2 + r_3 + 2(r_2 - r_1)\,\cn^2(Z;m)\,,
\label{e:KPcnoidal}
\ee
where 
$\cn(Z;m)$ is the Jacobian elliptic cosine \cite{NIST2010},
$m = (r_2-r_1)/(r_3-r_1)$ is the elliptic parameter,
$Z = \sqrt{(r_3-r_1)/6}\,(x+q y - c t - x_o)$
and 
$c = \frac13(r_1+r_2+r_3) + q^2$.
If $r_1,\dots,r_3$ and $q$ are constant, 
Eq.~\eqref{e:KPcnoidal} is an exact solution of the KP equation
describing periodic traveling wave oscillations with peaks localized along the straight lines
$Z = 2n K(m)$, with $n$ integer,
where
$K(m)$ is the complete elliptic integral of the first kind.
As $r_2\to r_1^+$, $m\to0$ and the solution limits to vanishing-amplitude harmonic oscillations.
Conversely, as 
$r_2\to r_3^-$, $m\to1$ and the solution limits to a line soliton.
A slanted DSW is obtained when $q= \mathrm{const}$ and $r_1,r_2\&r_3$ are given by a solution of 
the Whitham modulation system for the KP equation \cite{RSPA2017v473p20160695,supplement}:
$r_1(x,y,t) \equiv u_\min$,
$r_3(x,y,t) \equiv u_\max$
and 
$r_2(x,y,t) = R_2(\xi)$ with $\xi = (x+qy)/t - q^2$, where
$R_2(\xi) = u_\min$ for $\xi < \xi_\min$ and 
$R_2(\xi) = u_\max$ for $\xi > \xi_\max$,
with 
$\xi_\min = - \Delta + u_\min$, 
$\xi_\max = \frac23\Delta + u_\min$
and $\Delta = (u_\max - u_\min)$.
For $\xi_\min<\xi<\xi_\max$, $R_2(\xi)$ is given by inverting the 
above expression for $m$ as a function of $r_1,r_2\&r_3$, i.e., 
$R_2 = r_1 + m\,\Delta$, 
with 
$\xi = V_2(m)$
where
$V_2(m) = u_\max + \frac13\Delta (1+m + 2m(1-m)K(m)/(E(m)-(1-m)K(m)))$
and $E(m)$ is the complete elliptic integral of the second kind.
The above solution generalizes to 2D the well-known Gurevich-Pitaevski solution of the Whitham equations for the KdV equation \cite{JETP38p291}.
At the leading-edge of the DSW, $m\to1$, and the solution limits to a line soliton with amplitude $2\Delta$.
Conversely, at the trailing edge of the DSW, $m\to0$, and the solution limits to a constant.
All these solutions can also be obtained as generalizations of $y$-independent solutions 
through the so-called ``pseudo-rotation'' symmetry of the KP equation
\cite{SAPM73p183,RSPA2017v473p20160695}.
Namely, if $u(x,y,t)$ is any solution of Eq.~\eqref{e:KP}, then so is
$u(x + qy - q^2 t,y - 2q t, t)$
for any $q\in\Real$.
These solutions, along with slanted rarefaction waves, 
comprise the elemental building blocks that appear in the more complicated dynamical scenarios discussed below.

\medskip
\textbf{Wedge-shaped initial profiles.}
Below we show that a variety of dynamical outcomes can arise from sectorial Riemann problems, 
and we classify and quantitatively characterize the resulting novel nonlinear phenonema,
which include the Mach reflection and expansion of DSWs in two spatial dimensions.
Specifically, we study the behavior of solutions of Eq.~\eqref{e:KP}
arising from the following class of ICs:
\vspace*{-0.2ex}
\be
u(x,y,0) = U(x - X(y))\,,
\label{e:IC}
\ee
where $U(\xi)$ represents either a step up or a step down located at $\xi=0$.
Without loss of generality we can take either of two values 0 and 1 to the left and the right of the step, 
thanks to the scaling and Galilean invariances of Eq.~\eqref{e:KP}
\cite{RSPA2017v473p20160695}.
The curve $x = X(y)$ describes the transition profile in the $xy$ plane.
If $X(y)$ is constant, Eq.~\eqref{e:KP} reduces to the KdV equation, whose behavior is well 
characterized \cite{Whitham1974,JETP38p291,PHYSD333p11}.
Similarly, if $X(y) = q_oy$, with $q_o=\const$, one can take advantage of the pseudo-rotation symmetry 
of the KP equation
to see that the resulting behavior is simply a slanted rarefaction wave (for an upward step) or a slanted DSW (downward step).
The constant parameter $q_o$ quantifies the slope of the transition line in the $xy$-plane.

When the spatial profile of the step is nontrivial, novel two-dimensional wave structures arise.
In this Letter we study the time evolution of wedge-shaped profiles.  
Namely, we take $X(y)$ to be a smoothed-out version of   
$X(y) =  - q_o\,|y|$, again with $q_o = \const$.
Letting $u_\pm = \lim_{x\to\pm\infty} u(x,y,0)$,
the following four types of wedge~ICs arise:
(I)~downward step ($u_- = 1$ and $u_+=0$) with an acute angle ($q_o>0$);
(II)~downward step with an obtuse angle ($q_o<0$);
(III)~upward step ($u_- = 0$ and $u_+=1$) with an acute angle;
(IV)~upward step with an obtuse angle.
The time evolution of wedge types III and~IV produces two-dimensional rarefaction waves
\cite{supplement}.
Here we focus instead on the time evolution of wedge types I and~II,
which lead to wavebreaking and the generation of multidimensional DSWs.

Since Eq.~\eqref{e:KP} and the wedge ICs are invariant under $y \mapsto -y$, 
all solutions satisfy $u(x,y,t) = u(x,-y,t)$.  
Thus, $u_y(x,0,t) = 0$, which is the slip wall boundary condition at $y = 0$
for inviscid water/internal wave theory
\cite{baines_topographic_1998}.  Consequently, solutions constrained
to $y > 0$ correspond to an oblique step in amplitude incident with a
wall at $y = 0$ and the subsequent reflection ($q_o < 0$) or expansion
($q_o > 0$) of the flow.  (E.g., such scenarios could occur respectively
during the flood or ebb tide.)

\begin{figure*}[t!]
\centerline{%
\includegraphics[height=0.25\textwidth,trim=40 0 60 0,clip]{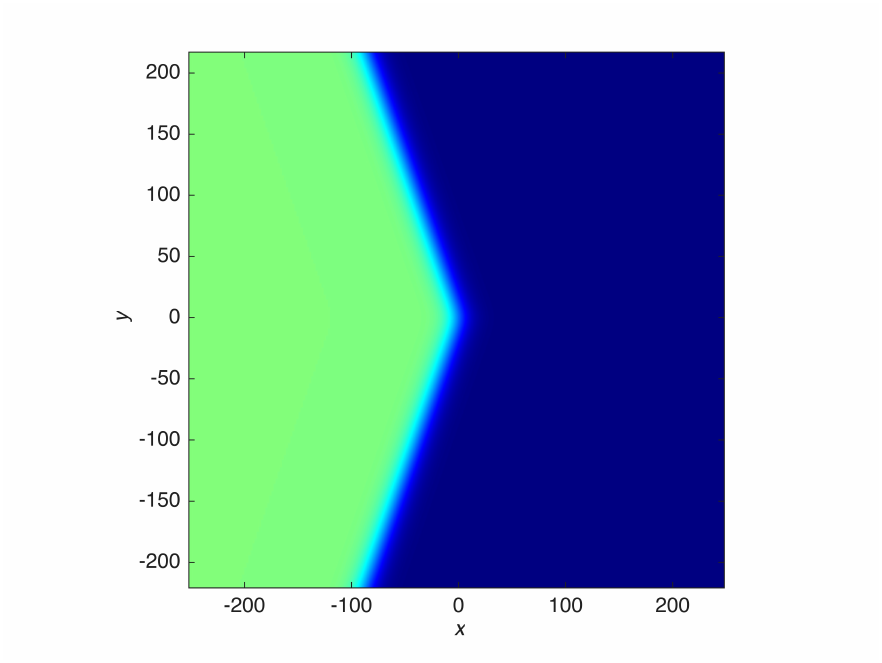}%
\includegraphics[height=0.25\textwidth,trim=65 0 60 0,clip]{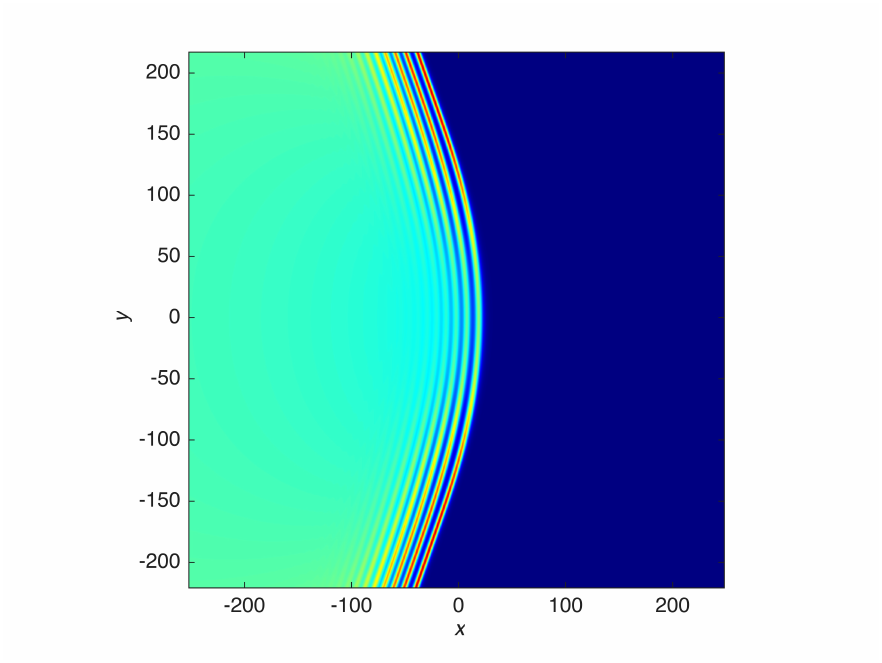}%
\includegraphics[height=0.25\textwidth,trim=65 0 60 0,clip]{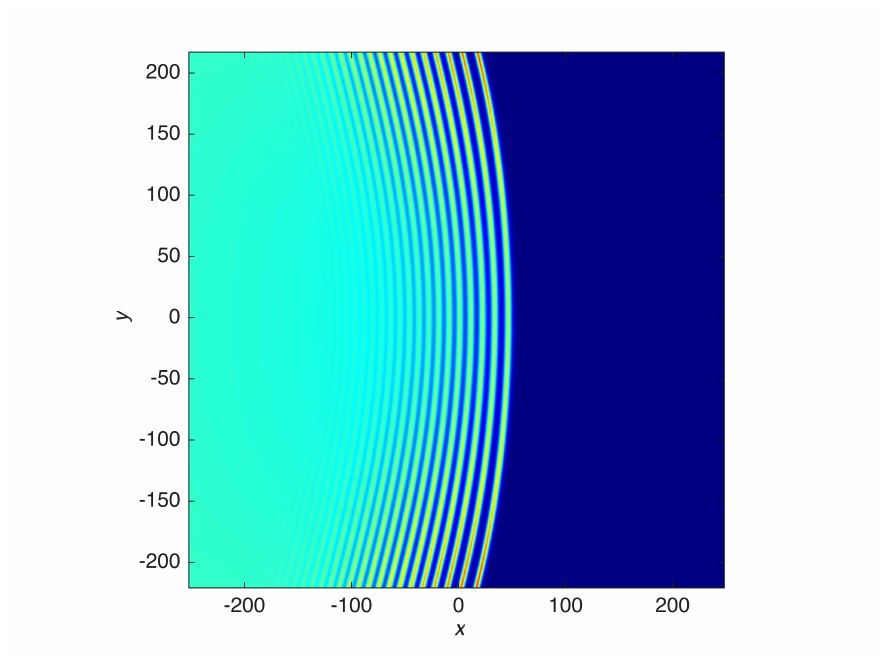}%
\includegraphics[height=0.25\textwidth,trim=45 0 50 0,clip]{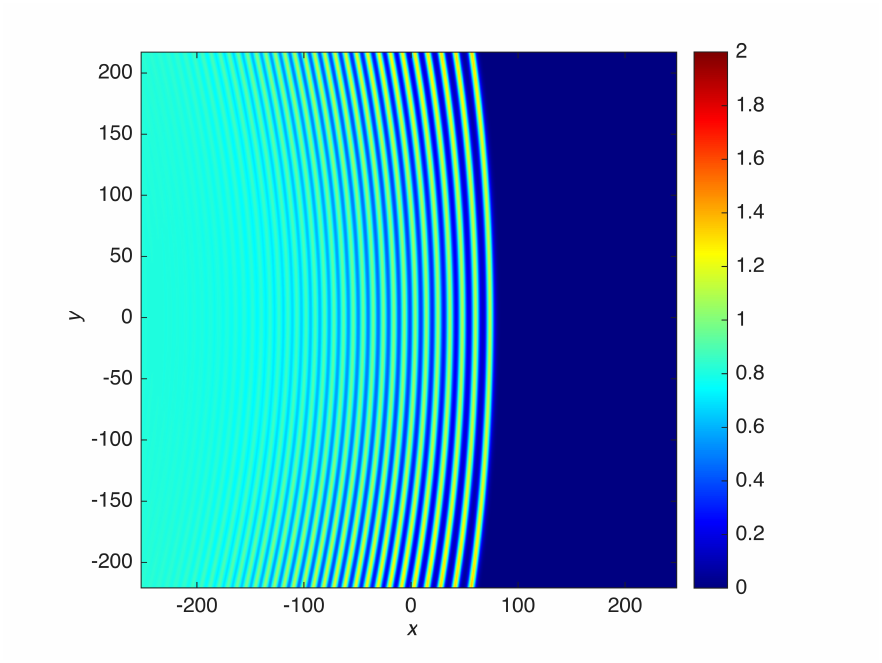}%
}
\vglue-\medskipamount
\caption{Solution of the KP equation produced by a subcritical type I wedge with $q_o = 0.4$ at times $t=0$, 75, 150 and 225.
Note the formation of a vertical DSW in the portion of the domain near the $x$-axis.}
\label{f:W1sub}
\medskip
\centerline{%
\includegraphics[height=0.25\textwidth,trim=40 0 60 0,clip]{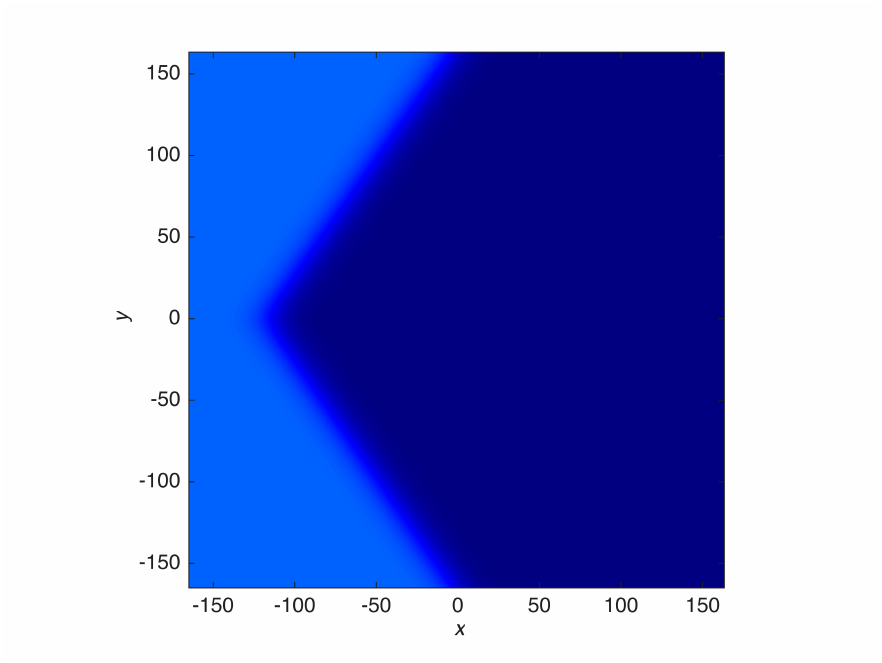}%
\includegraphics[height=0.25\textwidth,trim=65 0 60 0,clip]{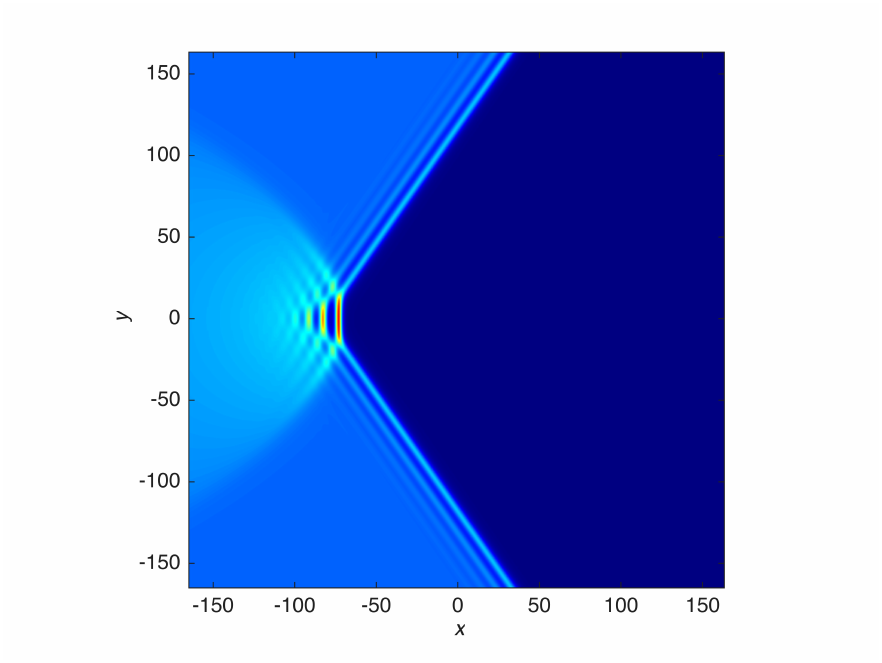}%
\includegraphics[height=0.25\textwidth,trim=65 0 60 0,clip]{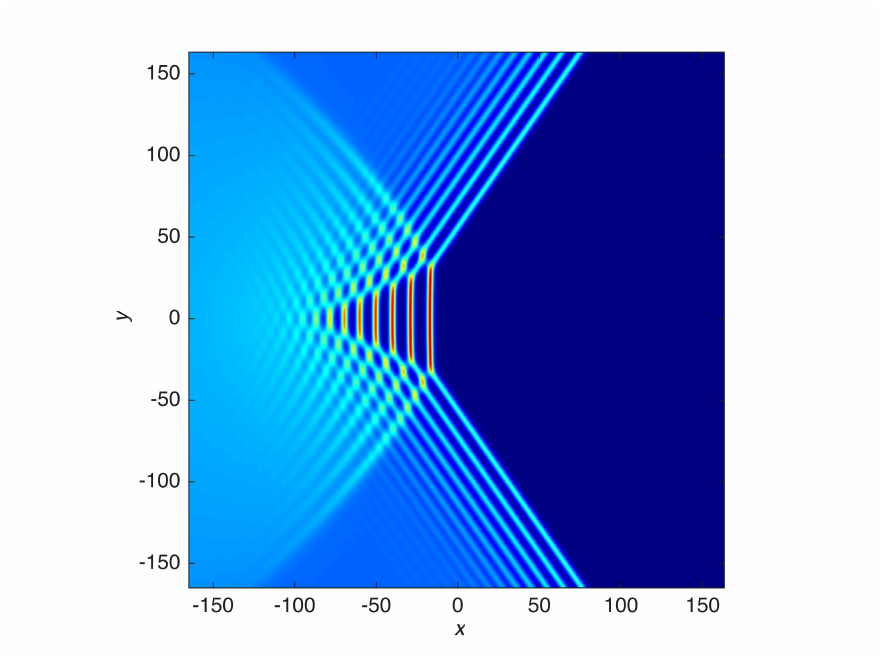}%
\includegraphics[height=0.25\textwidth,trim=45 0 50 0,clip]{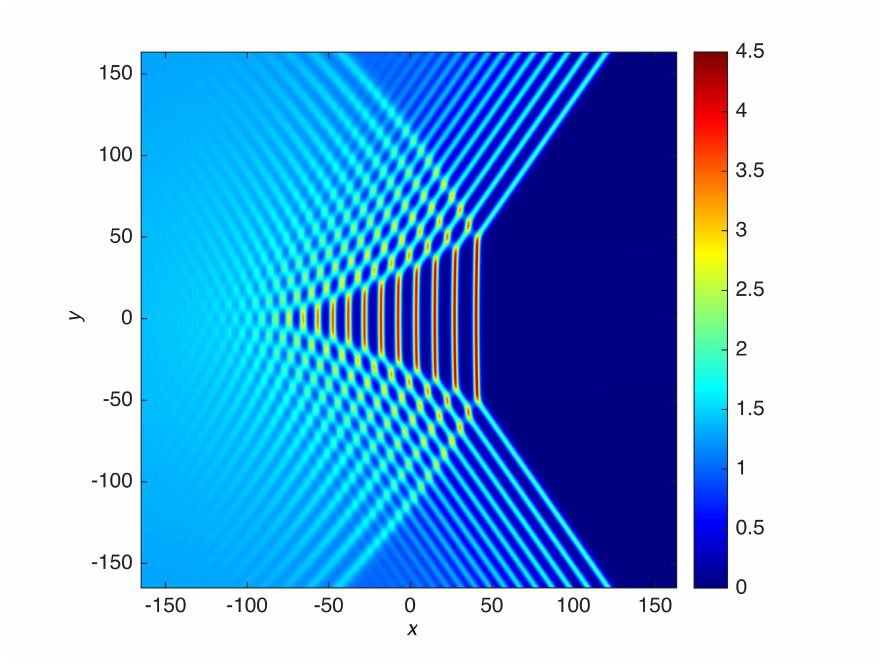}%
}
\vglue-\medskipamount
\caption{Solution of the KP equation produced by a subcritical type II wedge with $q_o = -0.7$ at times $t=0$, 40, 80 and 120.
Note the formation of a vertical DSW with high amplitude near the $x$-axis as well as
the presence of two expanding regions, symmetrically located with respect to the $x$-axis, 
characterized by multi-phase oscillations.}
\label{f:W2sub}
\medskip
\centerline{%
\includegraphics[height=0.25\textwidth,trim=40 0 60 0,clip]{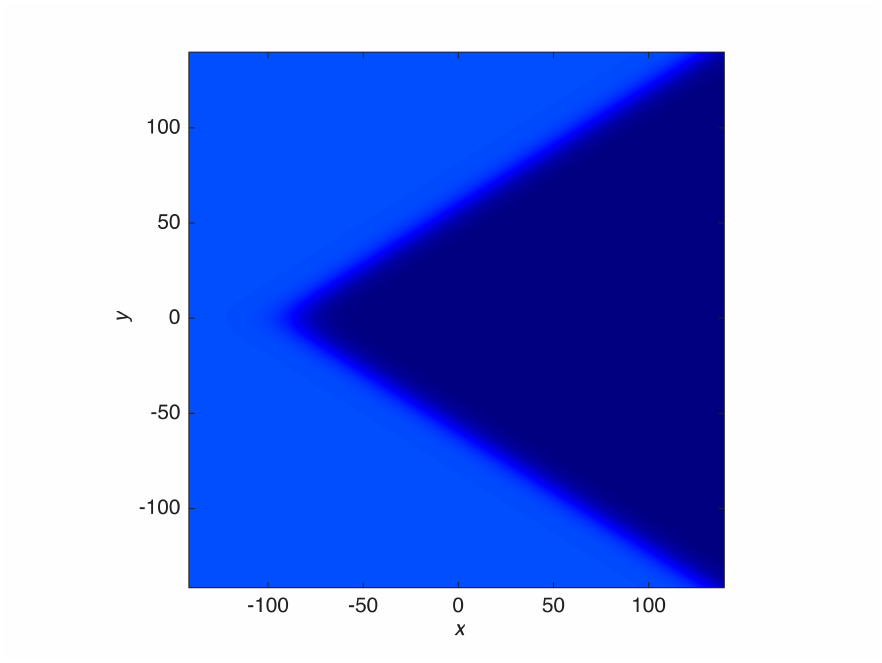}%
\includegraphics[height=0.25\textwidth,trim=65 0 60 0,clip]{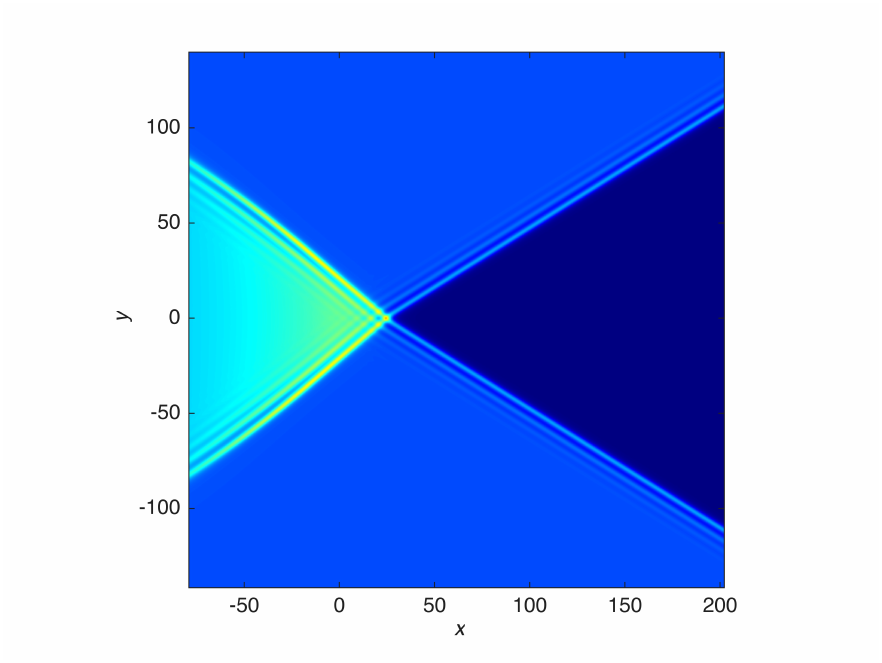}%
\includegraphics[height=0.25\textwidth,trim=65 0 60 0,clip]{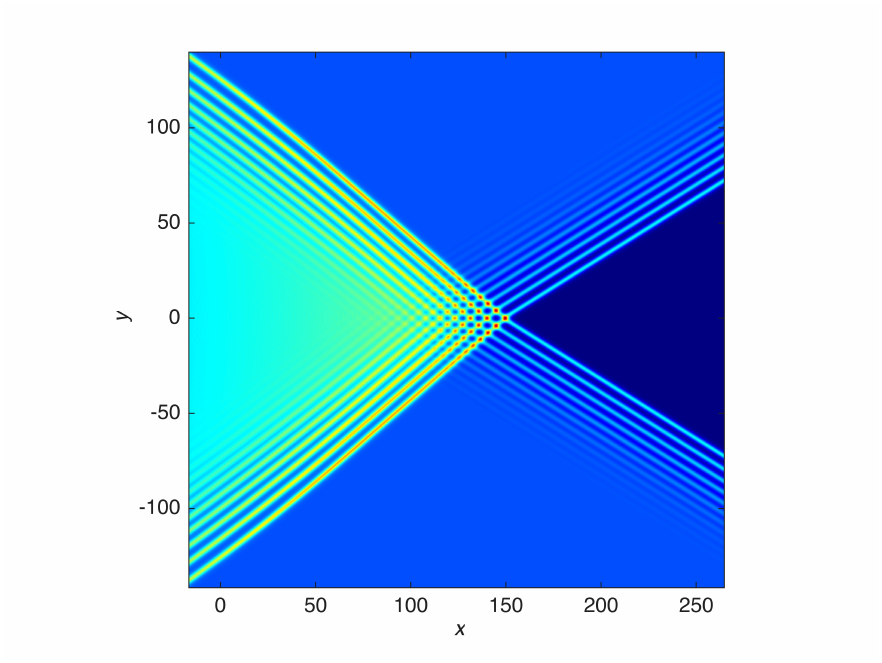}%
\includegraphics[height=0.25\textwidth,trim=45 0 50 0,clip]{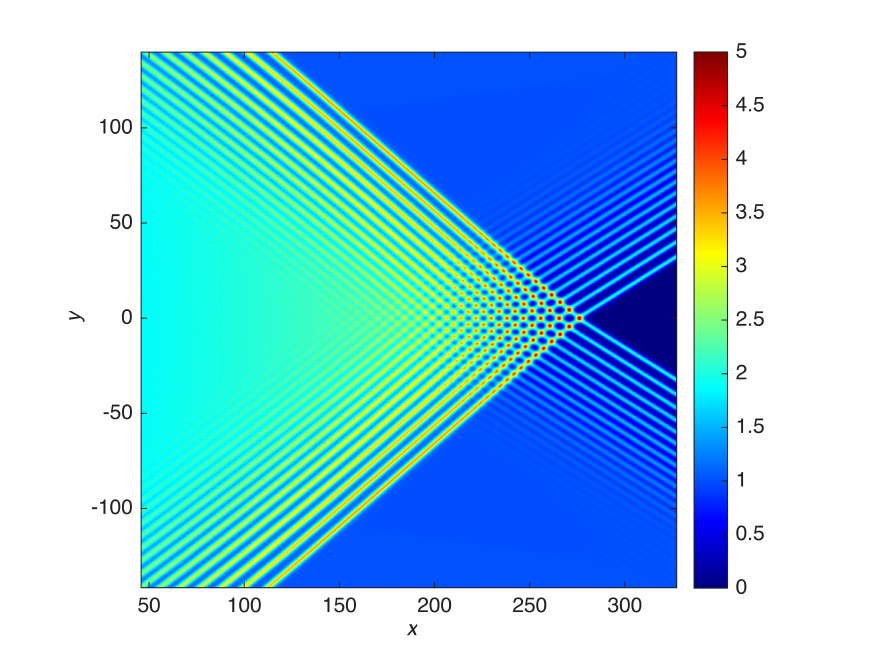}%
}
\vglue-\medskipamount
\caption{Same as Fig.~\ref{f:W2sub}, but for a supercritical type II wedge with $q_o = - 1.6$.
Note the formation of a single expanding region of multi-phase oscillations around the $x$-axis
as well as higher propagation speed of the interaction region along the $x$ direction compared to Fig.~\ref{f:W2sub}.
}
\label{f:W2super}
\end{figure*}

\medskip
\textbf{Temporal dynamics and its classification.}
The dynamics produced by wedge types I and~II are illustrated in 
Figs.~\ref{f:W1sub}, \ref{f:W2sub} and~\ref{f:W2super}
(plus Fig.~\ref{f:W1super} in the supplement \cite{supplement}).
These figures clearly demonstrate that the time evolution gives rise to a rich phenomenology of novel nonlinear phenomena,
including the emergence of curved DSW fronts and, 
in some cases, Mach stems and multi-phase regions with curved boundaries.
Below we demonstrate that, for both type I and~II wedges, 
a threshold $q_\crit = \sqrt2$ exists that discriminates between different dynamical outcomes.
Accordingly, \textit{we refer to wedges with $0< |q_o| < q_\crit$ as subcritical 
and wedges with $|q_o|> q_\crit$ as supercritical.}
For a generic initial jump with $u_- > u_+$,
the scaling invariance of the KP equation can be used to show that
the corresponding critical angle is $q_\crit = \sqrt{2\Delta}$,
with $\Delta = u_--u_+$.  

For both type~I and~II wedges, 
and irrespective of the value of $q$, 
the time evolution 
results in the formation of two slanted DSWs at large values of $|y|$,
which are accurately described by the modulation solutions discussed earlier.
The leading edge of each of these DSWs is a slanted line soliton
whose slope equals the slope $q_o$ of the initial datum, 
and whose amplitude $a_s=2$ equals twice the size of the initial jump,
as in the 1D case.
The crucial difference between subcritical and supercritical wedges lies in the behavior of the solution near $y=0$, as discussed next.%

\vspace*{-0.5ex}
One way to understand the subcritical/supercritical dichotomy is to look at the interaction between the leading-edge solitons in the slanted DSWs.
Equivalently, one can look at the interaction between soliton stems with amplitude $a_s=2$ and slopes $\pm q_o$,
similarly to \cite{JFM2021v909pA24,RSPA2022v478v20210823}.
Since the amplitude of the leading-edge soliton for the two slanted DSWs is $a_s=2$, 
the threshold $q_\crit = \sqrt2$ between subcritical and supercritical wedges 
is precisely the same as the threshold between ordinary and resonant two-soliton interactions \cite{PRL99p064103}. 

\vspace*{-0.3ex}
For subcritical type~I wedges, 
the time evolution produces, asymptotically in time, a region characterized by a vertical DSW, 
whose leading edge has constant in time amplitude and speed. 
In contrast, no such vertical DSW is produced in supercritical type~I wedges, 
and in this case the amplitude of the whole DSW around $y=0$ decays in time,
while its spatial profile tends to a parabolic shape (see later for details).

\vspace*{-0.5ex}
The dynamics produced by type~II wedges is even richer.
In this case the ICs are of compressive type 
as opposed to expansive,
which causes the two slanted DSWs to propagate into each other and interact.
As shown in Fig.~\ref{f:W2sub},
for subcritical type II wedges, the interaction between the two slanted DSWs 
produces an expanding region around $y=0$
characterized by a third, vertical DSW.
This phenomenon is a generalization to DSWs of the Mach stem produced by the refraction of two line solitons 
\cite{JFM79p171,JFM672p326,JPA36p10519,PRL99p064103}
or the Mach reflection of viscous shock waves \cite{mach_1875,krehl_discovery_1991,brio_mach_1992}.
Note the amplitude of the emergent DSW along $y=0$ is much larger than that of the slanted DSWs surrounding it.
In contrast, for supercritical type~II wedges no vertical stem is produced (cf.\ Fig.~\ref{f:W2super}).
Moreover, in both subcritical and supercritical type~II wedges, the interaction between these slanted DSWs produces expanding regions characterized by modulated multi-phase oscillations.
For subcritical type~II wedges, two such regions are produced, which are symmetrically located above and below the $x$-axis (cf.\ Fig~\ref{f:W2sub}).
In contrast, for supercritical type~II wedges, a single such region is produced, 
surrounding a portion of the $x$-axis  (cf.\ Fig~\ref{f:W2super}).

\begin{figure}[t!]
\kern\medskipamount
\centerline{%
\hglue-1em\includegraphics[width = 0.240\textwidth]{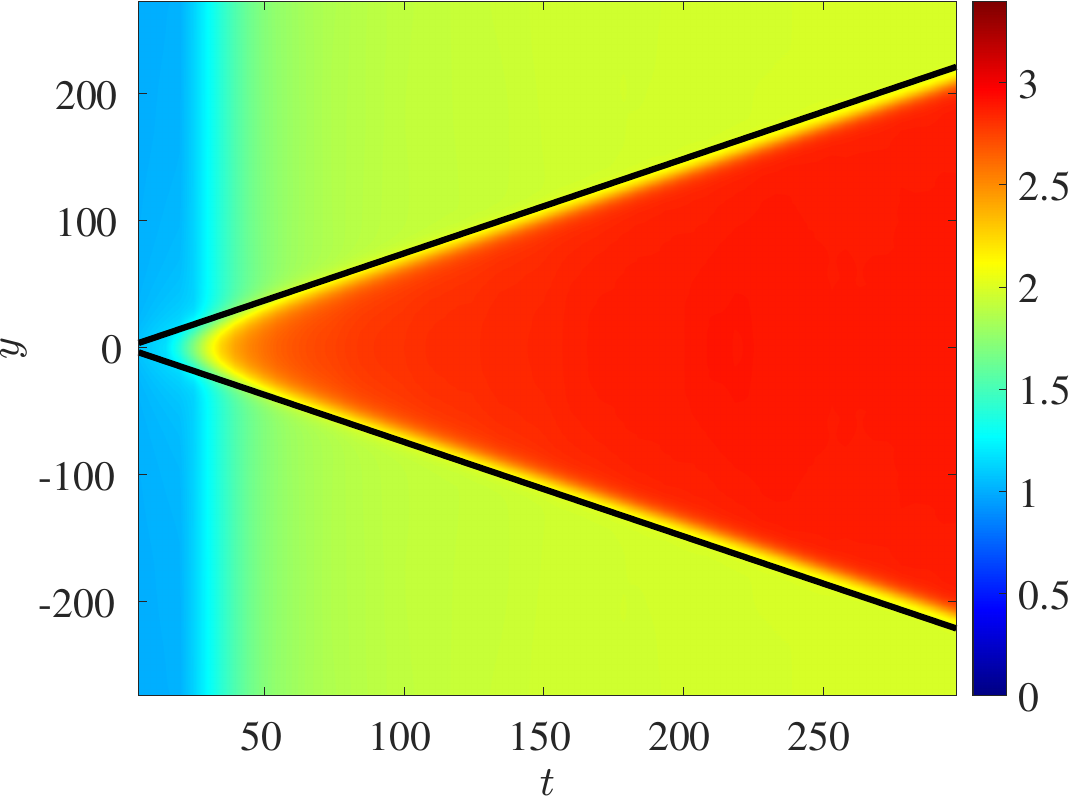}%
\includegraphics[width = 0.250\textwidth]{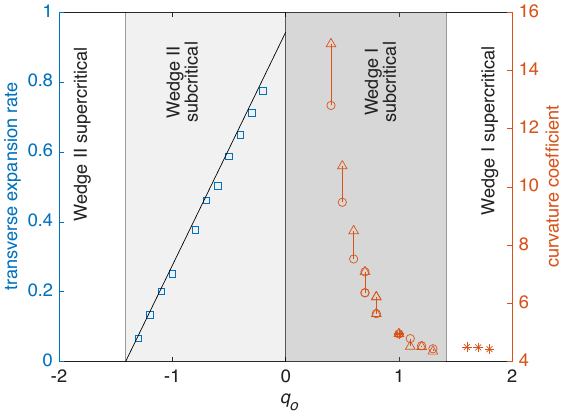}}
\kern-\smallskipamount
\caption{Left: 
Density plot of $u(x(t),y,t)$ showing the expansion of the leading-edge vertical soliton stem located at $x=x(t)$ in a subcritical type~II wedge with $q_o= - 0.3$, plus the analytical prediction from Eq.~\eqref{e:expansionrate} (black lines).
Right, $q_o<0$: 
Numerically computed transverse expansion rate $dy/dt$ of the vertical soliton stem in subcritical type~II wedges as a function of $q_o$ (blue squares) together with the theoretical prediction from Eq.~\eqref{e:expansionrate} (black line). 
Right, $q_o>0$:
Numerically computed temporal coefficient of the $y^2$ term in the parabolic front for supercritical type~I wedges (orange asterisks)
and subcritical type~I wedges in the time range $t\in[78,125]$ (orange circles) 
and $t\in[125,172]$ (orange triangles).
}
\label{f:expansion}
\vskip\bigskipamount
\centerline{\includegraphics[width=0.232\textwidth]{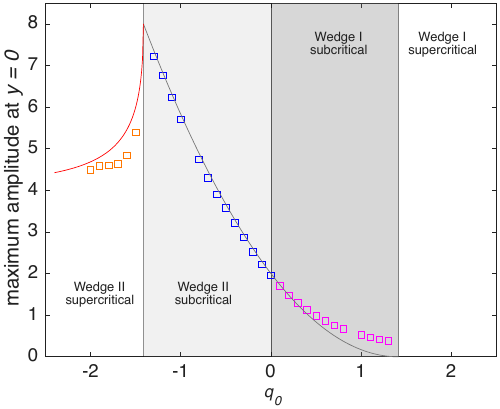}
\includegraphics[width=0.242\textwidth]{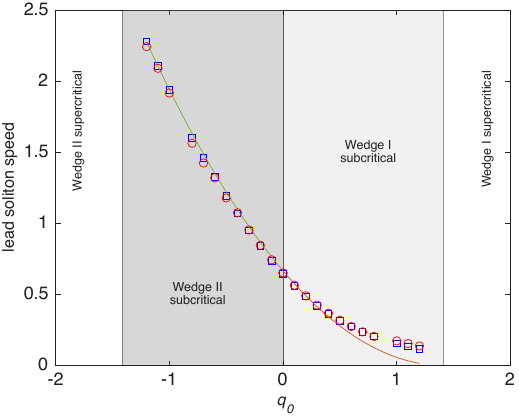}}
\kern-\smallskipamount
\caption{Left: Numerically computed amplitude (red and magenta squares) of the leading-edge vertical soliton stem for subcritical type~I and~II wedges as a function of $q_o$; maximum amplitude for supercritical type~II wedges (orange squares) and theoretical predictions from Eqs.~\eqref{e:solitonamplitude} and~\eqref{e:maxamplitude}.
Right: Numerically computed horizontal propagation speed $dx/dt$ for subcritical type~I and~II wedges
(squares), together with the theoretical prediction $C(A_o,0,0)$ (solid curve).}
\label{f:amplitude}
\kern-\medskipamount
\end{figure}

\medskip
\textbf{Quantitative analysis.}
Both subcritical and supercritical behaviors can be characterized quantitatively,
as discussed next. 
In particular, both the amplitude and the expansion rate of the soliton stem at the leading edge of the DSW 
in the central portion of the $y$ domain in subcritical type~II wedges 
are modeled by the dynamics of bent solitons~\cite{JFM2021v909pA24}.
Specifically, the central soliton stem in subcritical type~II wedges is located in the region $|y|\le V_o t$
(cf.\ Fig.~\ref{f:expansion}a),
where the vertical expansion rate $dy/dt = V_o(a,q)$ is given by
\vspace*{-0.8ex}
\be
V_o(a,q) = 2(\sqrt{a} + q)/3\,.
\label{e:expansionrate}
\ee
Here $q = q_o$, and $a = a_s$ is the amplitude of the leading-edge soliton in the slanted DSWs (i.e., $a_s = 2$ here).
As shown in Fig.~\ref{f:expansion}b, Eq.~\eqref{e:expansionrate} is in excellent agreement with the numerical results for all $-q_\crit<q_o<0$.
Similarly, the amplitude $A_o$ of the vertical stem in subcritical type~II wedges is given by 
\cite{RSPA2022v478v20210823}%
\vspace*{-1.4ex}
\be
A_o(a,q) = \big(\sqrt{a} - q\big)^2\,,
\label{e:solitonamplitude}
\ee
where again $q=q_o$ and $a = a_s$.
This is exactly the same prediction as for the amplitude of the central region in a bent-stem soliton interaction \cite{JFM2021v909pA24}.
As shown in Fig.~\ref{f:amplitude}a, 
this prediction is also in excellent agreement with the results of numerical simulations for all $ - q_\crit < q_o < 0$.
Finally, the numerically computed horizontal soliton speed $dx/dt$ of the leading-edge vertical soliton stem
also agrees very well with the soliton amplitude-speed relation $C(a,q,\bar u)$ with $a = A_o$ and $\bar u = q = 0$
(cf.\ Fig.~\ref{f:amplitude}b).
The largest solution amplitude is produced when $q_o = - \sqrt{a_s}$,
which yields $A_\mathrm{max} = 4a_s$,
the well-known four-fold amplification of an oblique soliton incident upon a wall.  
Since $a_s = 2\Delta$ for the slanted DSW, however, the final result is 
$A_\mathrm{max} = 8\Delta$,
an eightfold increase in the amplitude compared to the size of the jump.

While no vertical DSW is produced in supercritical type~II wedges,
the maximum solution amplitude at $y=0$ is approximately given by the maximum of the 2-soliton solution associated with two solitons with amplitude $a = a_s$ and slope $q = \pm q_o$ \cite{Kodama2017}, which is easily found to be 
\vspace*{-0.6ex}
\be
u_\mathrm{max}(q) = -2q\bigg(\sqrt{\sqrt2 q+2}\sqrt{2\sqrt2q-4}-2q\bigg).
\label{e:maxamplitude}
\ee
This prediction agrees fairly well with the results of numerical simulations for 
$q_o< - q_\crit$ (cf.\ Fig.~\ref{f:amplitude}a), and it also agrees with Eq.~\eqref{e:solitonamplitude} in the limit $q_o\to - q_\crit$.

The amplitude of the leading-edge vertical soliton stem in subcritical type~I wedges also agrees with Eq.~\eqref{e:solitonamplitude}
(cf.~Fig.~\ref{f:amplitude}a,
with small discrepancies when $q_o$ is close to $q_\crit$, 
likely due to the fact that measuring very small amplitudes is more sensitive 
to small numerical errors due to dispersive radiation).
The horizontal speed of the leading DSW front at $y=0$ compares well with the predicted speed $C(A_o,0,0)$ 
(cf.~Fig.~\ref{f:amplitude}b, again with small discrepancies for $q_o\sim q_\crit$).
In any case, for all $0<q_o<q_\crit$, the measured amplitude and speed agrees well with the soliton amplitude-speed relation given by $C(a,0,0)$.

In supercritical type~I wedges, 
the solution near $y=0$ tends to a DSW of the cylindrical KdV (cKdV) reduction of the KP equation, whose amplitude decays in time.
Recall that Eq.~\eqref{e:KP} admits a self-consistent reduction in which solutions depend on $x$ and $y$ only through the similarity variable $\xi = x + C(t)y^2$,
with $C(t) = c/(2+4t)$  
and $u(\xi,t)$ satisfies the cylindrical KdV equation,
$u_t + 6 u u_\xi + u_{\xi\xi\xi} + C(t)\,u = 0$
\cite{ablowitz_dispersive_2016,RSPA2017v473p20160695}.
By extracting the shape of the DSW near $y=0$, fitting it to a parabolic profile and extracting via a linear regression the time dependence of the curvature $C^{-1}(t)$, 
one can then measure how well a DSW approaches that of the cKdV equation. 
Indeed, Fig.~\ref{f:expansion}b shows that, for supercritical type~I wedges, the numerically obtained coefficient of $t$ in $C(t)$ 
reaches a steady value that agrees well with the prediction $4$ for the cKdV reduction.
Conversely, for subcritical type~I wedges, the temporal dependence is super-linear,
consistent with the fact that the solution near $y=0$ becomes a constant-amplitude vertical DSW.

\medskip
\textbf{Discussion.}
In summary, we have presented a scenario for the generation of two-dimensional DSWs, demonstrating the existence of a critical angle discriminating between regular reflection and Mach reflection of DSWs, and we have characterized the resulting solutions quantitatively.
The results of this work are robust with respect to numerical perturbations.
We expect that similar phenomena will arise in other related physical models, including the 2D-Benjamin-Ono\break equation, which governs the evolution of weakly 2D internal waves.

Applications for expansive ICs include the generation of parabolic DSWs from high-speed ferries in shallow water and atmospheric lee waves from mountains
\cite{lee_upstreamadvancing_1990,li_three-dimensional_2002,soomere_nonlinear_2007,mann_improve_2019}.
Conversely, compressive conditions model a flood-tide-generated undular bore obliquely impinging upon an obstacle, e.g., a sill, resulting in a Mach stem DSW (subcritical case) or a multi-phase
DSW (supercritical case) \cite{yuan_diffraction_2022}.

Multi-phase regions also arise in 1D interacting DSWs (e.g., the KdV equation).
A fundamental difference, however, is that in the 1D case the multi-phase regions are eventually extinguished \cite{CPAM55p1569}, in constrast to the present case, in which these regions expand in time.
Also notable is that the slope of each slanted DSW for supercritical type~I wedges changes as a result of the interaction with the other one, reminiscent of what happens for 1D and 2D soliton gases \cite{arxiv2408.05548}.

Avenues for future study include a characterization of the results via the KP-Whitham system (KPWS), namely the Whitham modulation equations for the KP equation~\cite{RSPA2017v473p20160695}.
This however is a nontrivial task, because the theory for (2+1)-dimensional systems of conservation laws \cite{Zheng2001} is still not as well developed.
To quantitatively describe the modulated multi-phase regions, one will need multi-phase
KP-Whitham equations.
Even though such equations can be obtained via the formalism of \cite{Krichever}, however, they have never been written down explicitly, let alone used.
Degenerate limits of such equations will also be needed to characterize the boundaries between one-phase and multi-phase regions, similarly to \cite{PHYSD236p44}.
Yet another direction will be a study of the quantitative behavior of type~III and~IV wedges, and in particular the question of whether there is also a critical angle in that case.
Finally, the ultimate goal will be the experimental realization and observation of these phenomena in 2D shallow/internal gravity water wave tanks or in the field.

\medskip
\textbf{Acknowledgments.}
We thank the Isaac Newton Institute for Mathematical Sciences for support and hospitality 
during the programme ``Dispersive Hydrodynamics'', when work on this paper was undertaken 
(EPSRC grant number EP/R014604/1).
This work was partially supported by the National Science Foundation under grant numbers DMS-2009487 (GB) and DMS-2306319 (MAH).

\input appendix_arxiv

\bibliographystyle{apsrev4-1}
\bibliography{allrefs}

\end{document}

%% file: appendix_arxiv.tex
\appendix
\section*{Appendix}
\setcounter{equation}0
\def\theequation{A.\arabic{equation}}

\textbf{Evolutionary form of the KP equation; constant-mass constraint.}
We begin by discussing a constraint that must be satisfied by the ICs for the KP equation in non-evolutionary form, namely Eq.~\eqref{e:KP}, 
which we rewrite here for completeness:
\vspace*{-0.6ex}
\be
(u_t +uu_x + u_{xxx} )_x + \sigma u_{yy} = 0\,,
\label{S:KP}
\ee
where again the sign $\sigma=\pm1$ distinguishes between the 
KPII and KPI equations, respectively.
Integrating~Eq.~\eqref{e:KP} in $x$ over the whole real axis yields
\be
\big[u_t + uu_x + u_{xxx} \big]_{x= -\infty} ^\infty = 
  - \sigma \int_\Real u_{yy}\d x \,.
\ee
If $u(x,y,t)$ tends to constant values as $x\to\pm\infty$,
the left-hand side of the above expression vanishes.
Interchanging the order of differentiation and integration in the right-hand side, 
for solutions that are bounded in $y$ one then obtains the constraint
\vspace*{-0.4ex}
\be
\frac{\partial}{\partial y}\int_\Real u(x,y,t)\,\d x = 0\,.
\label{S:constraint}
\ee
%
In all numerical simulations we will enforce this constraint by ensuring that
the integral of the IC with respect to $x$ is independent of $y$.
(See below for further details.)

\textbf{Evolutionary form of the KP equation.}
Note that the KP equation~\eqref{S:KP} can be written in evolutionary form as
\be
u_t +uu_x + u_{xxx} + \sigma \partial_x^{-1} u_{yy} = 0\,,
\ee
where the antiderivative $\partial_x^{-1}$ is defined as
\be
\partial_x^{-1} f(x) = \frac12 \bigg( \int_{-\infty}^x f(\xi)\,\d\xi - \int_x^\infty f(\xi)\,\d\xi \bigg)\,.
\ee
Equivalently, $\partial_x^{-1}$ can also be understood as the operator whose Fourier multiplier is the singular symbol $-i/k$,
as is done in the numerical method used to integrate Eq.~\eqref{S:KP}, which is discussed next.

\textbf{Numerical integration algorithm.} 
%
The KP equation~\eqref{S:KP} has several features that make it challenging to simulate numerically, such as  nonlinearity, nonlocality, and a stiff fourth derivative term.
The numerical algorithm chosen to integrate Eq.~\eqref{e:KP} in time is the fourth-order
integrating factor Runge-Kutta (IFRK4) method, as in~\cite{JNLSCI17p429}.
This method treats the nonlocal and stiff portions of the PDE with spectral accuracy, 
and the nonlinear portion of the equation with fourth order accuracy. 
In (2+1) dimensions the derivation is not trivial. 

We begin by defining the 2D Fourier transform as
\bse
\be
\^f(k,l) = \F[f(x,y)] = \iint_{\Real^2} e^{-i(kx+l y)}f(x,y)\,\d x \d y\,,
\label{e:FTdef}
\ee
which is inverted by
\be
f(x,y) = \F^{-1}[\^f(k,l)] = \frac{1}{2\pi}\iint_{\Real^2} e^{i(kx +l y)}F(k,l)\,\d k \d l\,.
\ee
\ese
The Fourier transform of Eq.~\eqref{e:KP} then yields
\be
\label{S:KPFT}
k\hat u_t = -k\F[uu_x] + ik^4\hat u - i\sigma l^2 \hat{u}\,.
\ee
It is therefore convenient to define
\be
\label{e:KP_u2v}
\hat v(k,l,t) := e^{-i(k^3 - \sigma l^2/k)t}\hat u(k,l,t)\,,
\ee
which solves the linear part of the ODE Eq.~\eqref{S:KPFT}
Substituting Eq.~\eqref{e:KP_u2v} into Eq.~\eqref{S:KPFT} yields
\be
\hat v_t  = - e^{-i(k^3 - \sigma l^2/k)t} \F[uu_x]\,,
\label{e:KP_vt}
\ee
which is an ODE without stiff terms, and which can therefore be efficiently solved with RK4. 
Once $\hat v$ is known, one can calculate $u$ by inverting Eq.~\eqref{e:KP_u2v}, namely:
\be
u(x,y,t) = \F^{-1}\big[e^{i(k^3 - \sigma l^2/k)t}\hat v(k,l,t)\big]\,.
\label{e:kp_uhat}
\ee
Summarizing, the algorithm behind the numerical integration can be described as follows:
\begin{enumerate}
\advance\itemsep-4pt
\item 
    Begin by constructing spatial and corresponding wavenumber grids.
\item 
    Take the FFT of the the IC evaluated and note that $\hat v(k,l,0) = \hat u(k,l,0)$.
\item
    At each time step, evolve Eq.~\eqref{e:KP_vt} in time using RK4.
    Note that, in order to do so, one needs to evaluate the right hand side of the PDE four separate times. 
    To this end, at each stage of the RK4 one must:
    \vspace*{-0.6ex}
    \begin{enumerate}
    \advance\itemsep-2pt
        \item 
        Use Eq.~\eqref{e:KP_u2v} to get $\hat u(k,l,t)$ from $\hat v(k,l,t)$ at the current time step;
        \item 
        Take the IFFT of $\hat u(k,l,t)$ to get $u(x,y,t)$;
        \item
        Take the IFFT of $ik\hat u(k,l,t)$ to get $u_x(x,y,t)$;
        \item
        Finally, take the FFT of the term $uu_x$ to get the RHS of Eq.~\eqref{e:KP_vt}.
    \end{enumerate}
\item 
At regularly spaced time intervals, use Eq.~\eqref{e:kp_uhat} to store a snapshot of the solution of the 
KP equation~\eqref{e:KP}.
\end{enumerate}
Equivalently, one can write $\F[u u_x] = \frac12 ik \F[u^2]$ in Eq.~\eqref{e:KP_vt}, which allows one to eliminate step~(c) above and evaluate the RHS of the equation with one less IFFT.

\textbf{Numerical zero-mass constraint.}
The implementation of the IFRK4 algorithm described above requires dealing with additional constraints related to Eq.~\eqref{S:constraint},
as we discuss next.

Recall that the exponential factor appearing in Eqs.~\eqref{e:KP_u2v} and~\eqref{e:kp_uhat} is $\exp[i(k^3 - \sigma l^2/k)t]$.
The division by $k$ is singular for the $k=0$ mode.
On the other hand, no singularity arises for $l=0$.
Similarly, 
Eq.~\eqref{S:KPFT} shows that no singularity arises if $\hat u(0,l,t)=0$.
The definition~\eqref{e:FTdef} of the Fourier transform implies that the condition $\hat u(0,l,t)=0$ corresponds to the zero-mass condition
\vspace*{-1ex}
\be
\int_{\Real} u(x,y,t)\,\d x = 0\,
\label{e:zeromass}
\ee
for all $y\in\Real$.
That is, the total mass of the solution in the $x$ direction should be zero
in order to be apply the above-described integration methods without introducing numerical singularities.
Note that, if Eq.~\eqref{e:zeromass} is satisfied for all $y\in\Real$, Eq.~\eqref{S:constraint} is also satisfied.
Note also that, when both $k=0$ and $l=0$, Eq.~\eqref{S:KPFT} is trivially satisfied.
On the other hand, if Eq.~\eqref{e:zeromass} is satisfied for all $y\in\Real$, one has that 
$\hat u(0,0,t) = \iint_{\Real^2} u(x,y,t)\,\d x\d y = 0$ for all $t$.

In the numerical implementation of the IFRK4 method, we deal with this constraint by manually setting $\^u(0,l,t) = 0$ at each time step 
for all nonzero $l$ 
and by setting $\^u(0,0,t)$ equal to its value at $t=0$.
Note that this constraint can be satisfied satisfied without loss of generality by using the Galilean invariance of the KP equation, namely, the fact that 
an overall upward/downward shift of $u(x,y,t)$ simply amounts to a constant shift in the velocity of the solution.

\begin{figure}[t!]
\kern\medskipamount
\centerline{\includegraphics[width=0.45\textwidth]{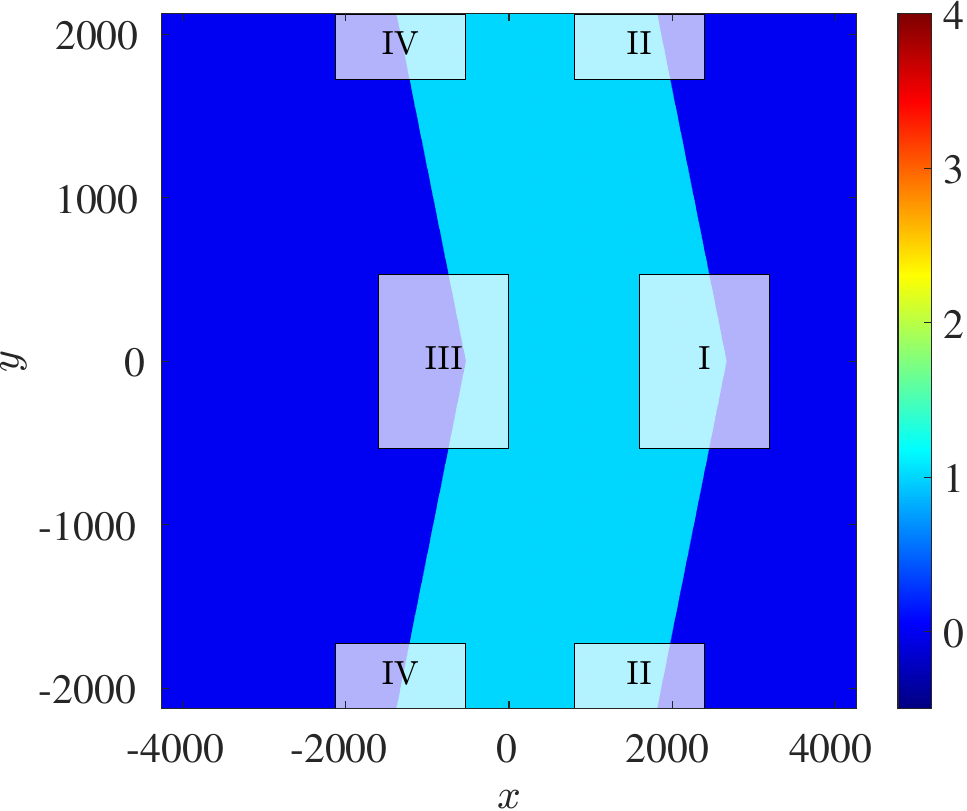}}
\caption{The four wedge-type ICs combined on a single, enlarged physical domain. 
Here $|q_o| = 0.4$.}
\label{f:combinedICs}
\end{figure}

\textbf{Combined spatial domain.}
The implementation of the numerical integration algorithm present
s multiple challenges. 
The first one is that, in order to use Fourier methods, the IC in the computational domain must be periodic in $x$ and $y$. 
One can deal with this constraint to by combining all four types of ICs into a single IC on an enlarged spatial domain,
as shown in Fig.~\ref{f:combinedICs}.
Rather than evolving each of the four wedge types separately, we then run a single simulation 
on the enlarged domain that contains all four wedge ICs. 
The time evolution of each individual wedge IC is then recovered by simply selecting a different portion of this enlarged domain (cf.\ Fig.~\ref{f:combinedICs}).

The size of the enlarged spatial domain must obviously be large enough so that effects from one wedge do not leak into any of the other subdomains. Therefore, the size of the spatial domain depends upon the desired $t_\text{max}$. 

\textbf{Numerical IC implementation.}
Next we discuss the numerical implementation of the ICs.
A~sharp step discontinuity features an immediate rise (or drop) between the values $0$ and $1$ for $u$ in the $x$~direction.
This corresponds to choosing an IC $u(x,y,0)$ in Eq.~\eqref{S:KP} as
$u(x,y,0) = H(x - x_1 + q|y|) - H(x - x_2 + q|y|)$,
where $H(x)$ is the Heaviside step function,
and where the quantities $x_1$ and $x_2$ vary the width and placement of the plateau,
and $q_o$ its slope in the $xy$-plane.
The width of the plateau is simply $x_2 - x_1$.
However such a choice of IC has the drawback that 
the Heaviside function and absolute value functions in $y$ are not smooth. 
This has the effect of exciting high-wavenumber Fourier modes that can spoil the accuracy of the results and
could even result in instabilities under time evolution. 
Therefore it is convenient to replace both $H(x)$ and $|y|$ with suitable smooth functions, as described next.

To minimize high-wavenumber dispersive components, we smooth out the transition between the values $0$ and $1$ along the $x$~direction by replacing the Heaviside step function with 
\vspace*{-0.6ex}
\be
H_\delta(\xi) = \frac12 \qty[1+\tanh(\delta\xi)]\,,
\ee
where the parameter $\delta$ quantifies the steepness of the transition.
If desired, the actual Heaviside function can be recovered in the limit $\delta\to0^+$.
Similarly, the corner at $y=0$ produced the absolute value function $|y|$ in the ICs is smoothed out by 
replacing it with $A(y)$, where $A'(y)$ is expressed in terms of $\tanh$ with a steepness parameter $a$, 
i.e., 
\vspace*{-0.4ex}
\be
A(y) = \int_0^y \tanh(ay)\,\d y = \frac1a \log(\cosh(ay))\,.
\nonumber
\ee
Since the IFRK4 code uses periodic boundary conditions in $y$, however,
the corner of the IC at $y=\pm y_\text{max}$ must also be smoothed out. 
This can be accomplished by using additional $\tanh$ functions, i.e., 
letting
\be
A'(y) = \tanh(ay) - \tanh(a(y+y_\text{max})) + \tanh(a(y-y_\text{max})) \label{e:multitanh}\,,
\ee
so that $A'(y)$ is also continuous at $y = \pm y_\text{max}$,
and where we used the fact that we will take $y_\mathrm{min} = - y_\mathrm{max}$.
Accordingly, we take
\vspace*{-1ex}
\begin{multline}
A(y) = \frac{1}{a} \big[ \log\cosh(ay) - \log\cosh(a(y+y_\mathrm{max})) \\
  - \log\cosh(a(y-y_\mathrm{max})) + 2\log\cosh(a y_\mathrm{max}) \big]\,,
\label{e:A}
\end{multline}
where the integration constant ensures that $A(0) = 0$.
Implementing this definition must also be done carefully, because the spatial domains are very large, 
and therefore the hyperbolic cosines becomes exponentially large.
The solution to this problem is to take advantage of the fact that, for all $z\in\Real$,
\be
\log\cosh z = |z| - \log2 + \log(1+e^{-2|z|})\,.
\label{e:logcosh}
\ee
The numerical implementation of $A(y)$ then uses Eq.~\eqref{e:A}
with all the $\log\cosh$ functions evaluated using Eq.~\eqref{e:logcosh}.

Summarizing, we implement the step-like wedge initial conditions I--IV on the combined spatial domain as 
\vspace*{-0.5ex}
\be
u(x,y,0) = H_\delta(x - x_1 + q_o A(y)) - H_\delta(x - x_2 + q_o A(y))\,,
\label{e:kp_wedge_initial}
\ee
with $H_\delta(x)$ and $A(y)$ as above,
where varying $\delta$ and $a$ changes the steepness of the step in the $x$ and $y$ directions, and varying $q_o$ adjusts the slope of the wedge.

\begin{figure*}[t!]
\centerline{%
\includegraphics[height=0.25\textwidth,trim=40 0 60 0,clip]{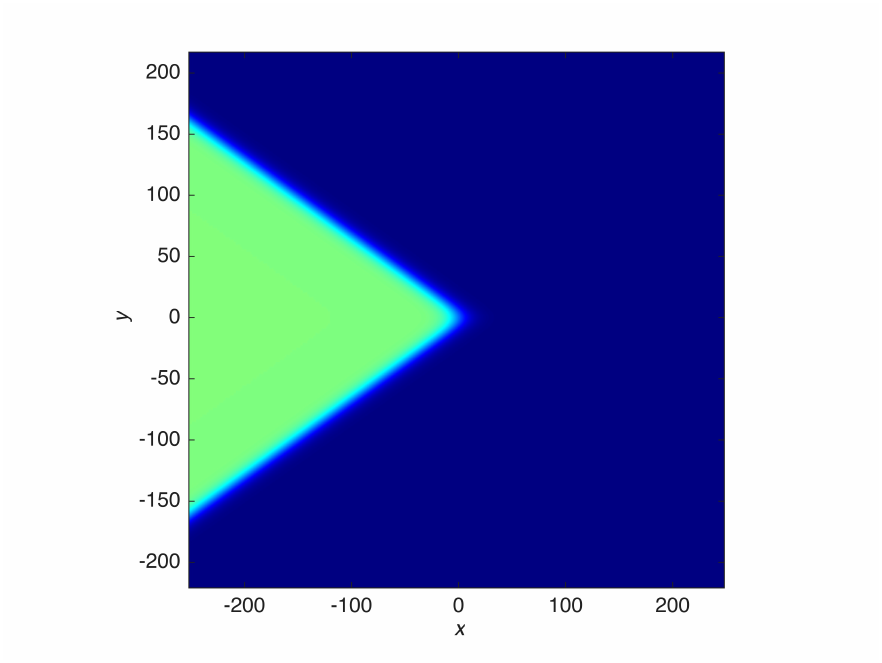}%
\includegraphics[height=0.25\textwidth,trim=65 0 60 0,clip]{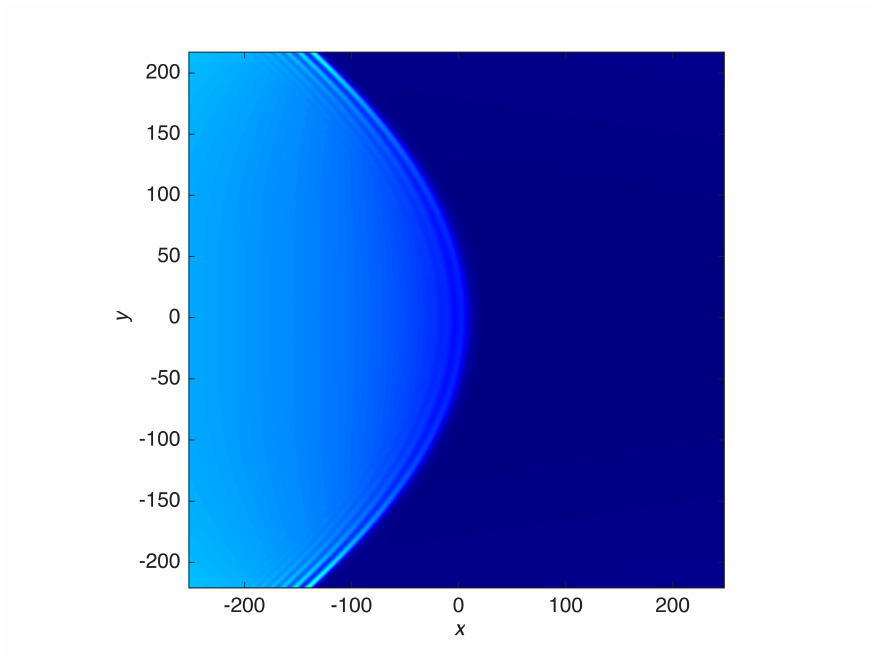}%
\includegraphics[height=0.25\textwidth,trim=65 0 60 0,clip]{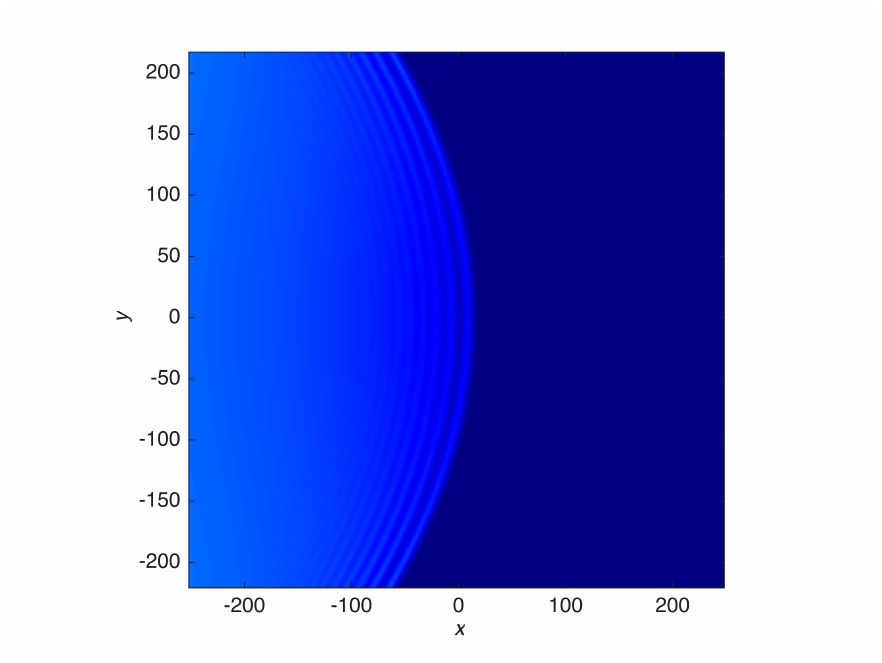}%
\includegraphics[height=0.25\textwidth,trim=45 0 50 0,clip]{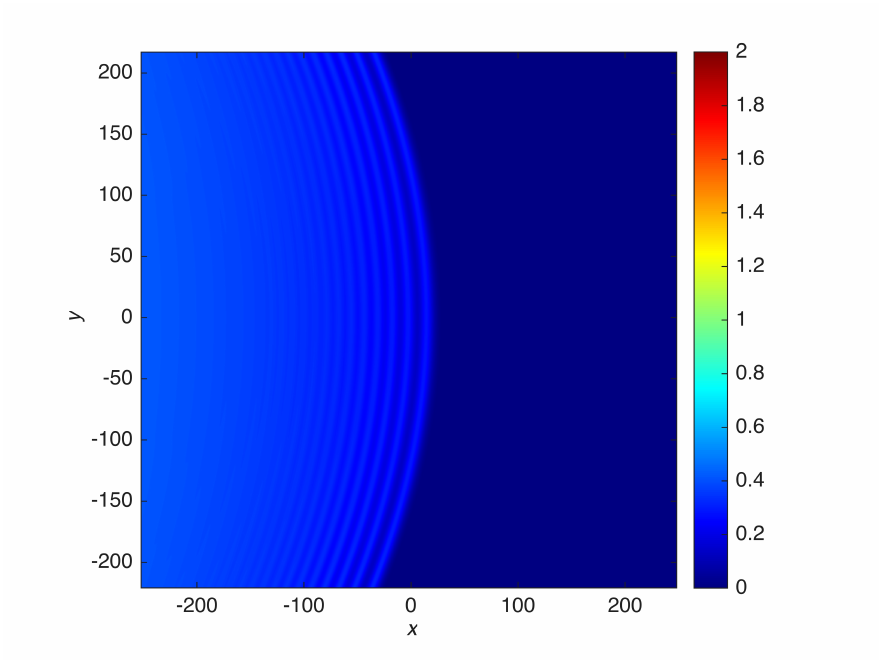}%
}
\vglue-\smallskipamount
\caption{Same as Fig.~\ref{f:W1sub}, but for a supercritical type I wedge with $q_o = 1.6$.
Note how the amplitude of oscillations near $y=0$  
and the corresponding propagation speed 
are much smaller compared to those in Fig.~\ref{f:W1sub}.}
\label{f:W1super}
\end{figure*}

\textbf{Artificial damping.} 
The wedge ICs discussed in the main text result in the production of trailing dispersive waves that travel in the negative $x$ direction. 
Since the numerical integration algorithm employs periodic boundary conditions, 
we want to prevent these phenomena from reaching the domain boundary and subsequently corrupting the nonlinear phenomena on the other side. 
Taking the $x$ domain to be large postpones the boundary crossing.  However, the size of the domain is ultimately constrained by numerical considerations.

An efficient solution to this issue is to use artificial damping,
generalizing the approach of \cite{LiuTrogdon} to 2 spatial dimensions 
by adding the term $\alpha\sigma(x)(u_{xx} + u_{yy})$ to the right hand side of the KP equation~\eqref{S:KP}, where the function $\sigma(x)$ defines the region in which damping will take place, and $\alpha$ quantifies the amount of damping applied. 
This dissipative term flattens out the trailing phenomena over the unused portion of the computational domain, such that any phenomena that crosses the periodic boundary are minimized.
Specifically, we take
\be
\sigma(x) = 1 - [\tanh(\delta(x-x_3)) - \tanh(\delta(x-x_4))]/2,
\ee
where $\delta$ is the same smoothing coefficient used for the initial conditions, and $x_3$ and $x_4$ are chosen so that the region of the spatial domain in which the DSWs are contained is not damped. 

To evolve the system with the artificial damping term added we use an alternating direction Crank-Nicholson implicit method. Therefore at each time step, the numerical solver performs the following steps:
(i) Evolve the system according to the KP equation, as described above.
(ii) Evolve the system according to damping term in the $x$ direction over the same time.
(iii) Evolve the damped portion of the domain over the same time step subject to damping in the $y$ direction.

The reason for using an implicit method is obviously the stiffness of the dissipative terms,
and the use of the alternating direction scheme allows one to use tri-diagonal matrices in the Crank-Nicholson scheme.  The resulting linear systems of equations are solved with a least squares solver. 
The inclusion of this solver roughly triples the run time. 
However, the artificial damping is effective enough at curtailing the dispersive phenomena that the overall size of the computational domain (and with it the number of grid points) can be decreased, 
resulting in a decrease in the overall simulation time for a given desired accuracy.

\textbf{Integration parameters.}
In order to capture the long-time behavior of the system, the computational domain must be large enough to ensure the simulation results remain accurate. 
As mentioned above, the primary concern is that phenomena generated by one wedge type eventually leak into the portion of the spatial domain associated with another wedge type 
(cf.\ Fig.~\ref{f:combinedICs}).
Along the $x$ direction,
we chose $x_\mathrm{xmax}$ and $x_\mathrm{xmin} = -x_\mathrm{max}$, with $x_\mathrm{max} = 5000$, resulting in a total domain size of 10,000, which ensured that any phenomena trailing wedges III and IV would not reach wedges I and II in the allotted time. 
Along the $y$ direction, the domain is similarly chosen to be symmetric with respect to $y=0$, so that $y_\mathrm{min} = - y_\mathrm{max}$.
The choice for the size of $y$ depends on the angle of the wedge, as determined by the parameter~$q_o$. 
The ratio between $x_\mathrm{max}$ and $y_\mathrm{max}$ must be chosen so that initial conditions 
do not leak across the different portions of the domain, irrespective of the wedge angle. 
The choice $y_\mathrm{max} = x_\mathrm{max}/4$ is found to provide sufficient separation for all values of $q_o$ considered.

The values of the parameters were chosen so that phenomena from wedge~I has not traveled far enough in the $y$ direction to impact wedge II, and vice-versa. 
Similarly, phenomena from wedge III and wedge IV have not leaked either. 

\begin{figure}[t!]
\kern-\medskipamount
\centerline{\includegraphics[height=0.25\textwidth,trim=40 0 40 0,clip]{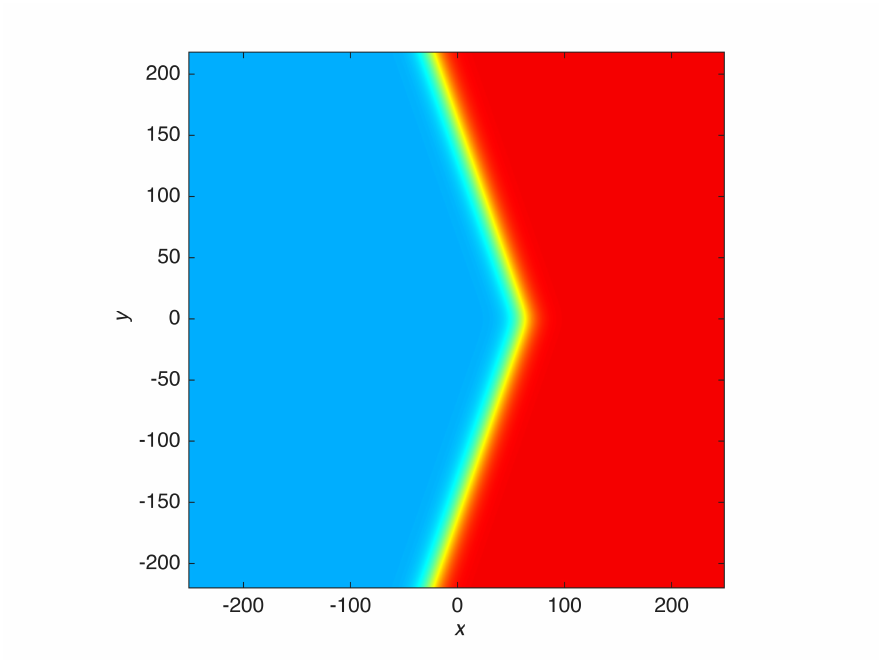}
\includegraphics[height=0.25\textwidth,trim=40 0 40 0,clip]{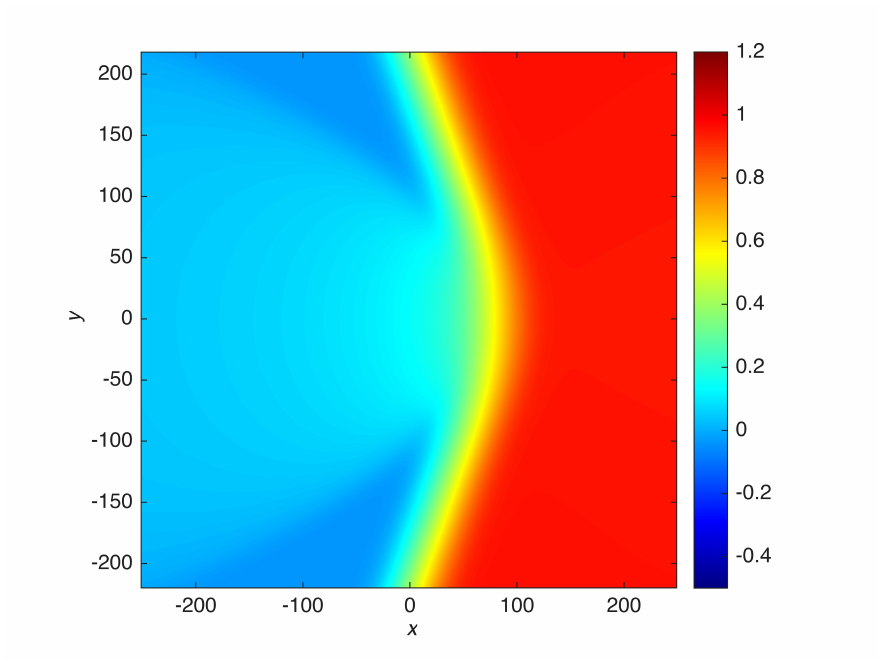}}
\kern-\medskipamount
\caption{Evolution of a type III wedge at time $t=0$ (left) and $t= 43$ (right) with $q_o = 0.4$, 
giving rise to two-dimensional rarefaction waves.}
\label{f:W3sub}
\vskip\medskipamount
\centerline{\includegraphics[height=0.25\textwidth,trim=40 0 40 0,clip]{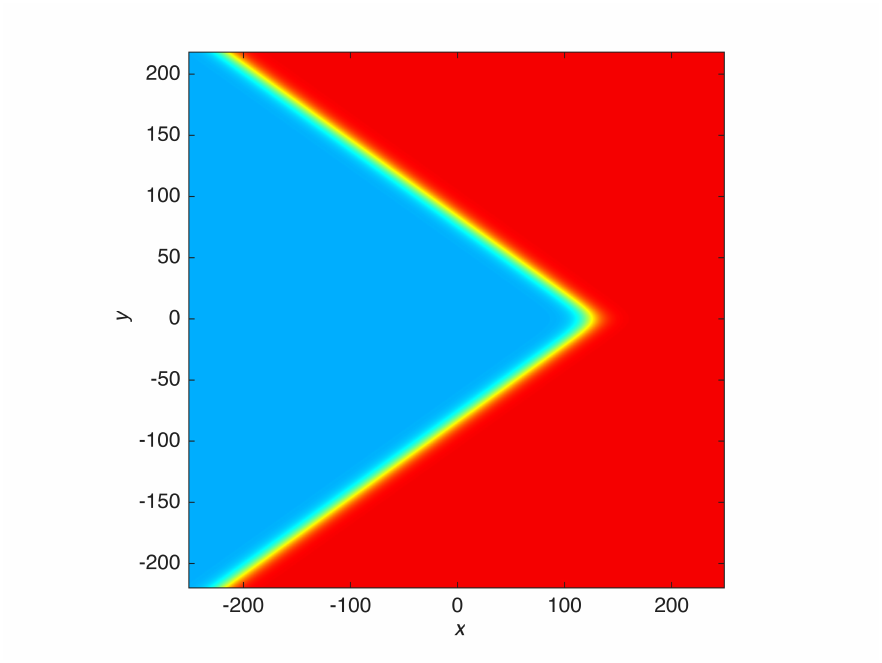}
\includegraphics[height=0.25\textwidth,trim=40 0 40 0,clip]{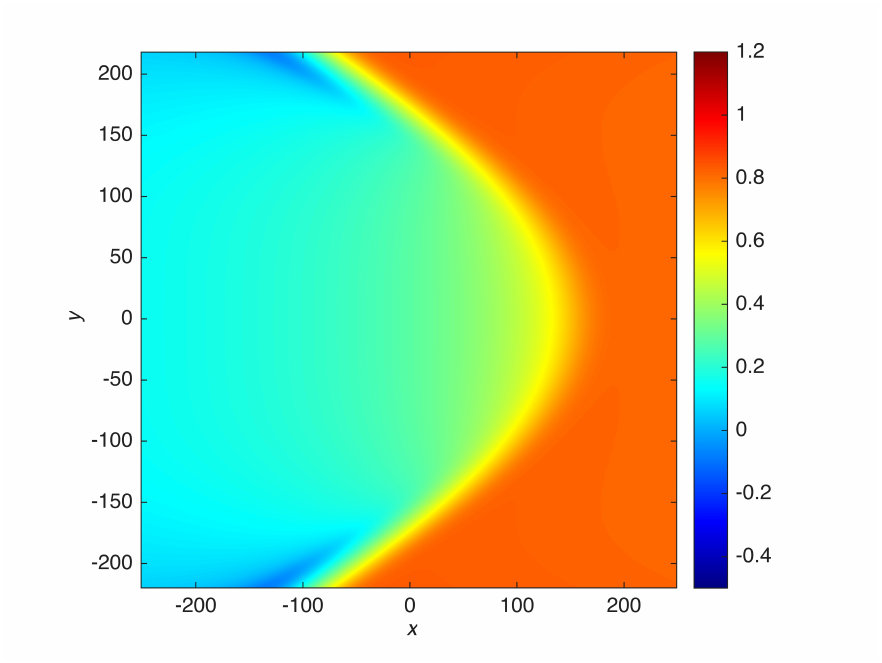}}
\kern-\medskipamount
\caption{Same as Fig.~\ref{f:W3sub}, but for $q_o = 1.6$.}
\label{f:W3super}
\vskip\medskipamount
\centerline{\includegraphics[height=0.25\textwidth,trim=40 0 40 0,clip]{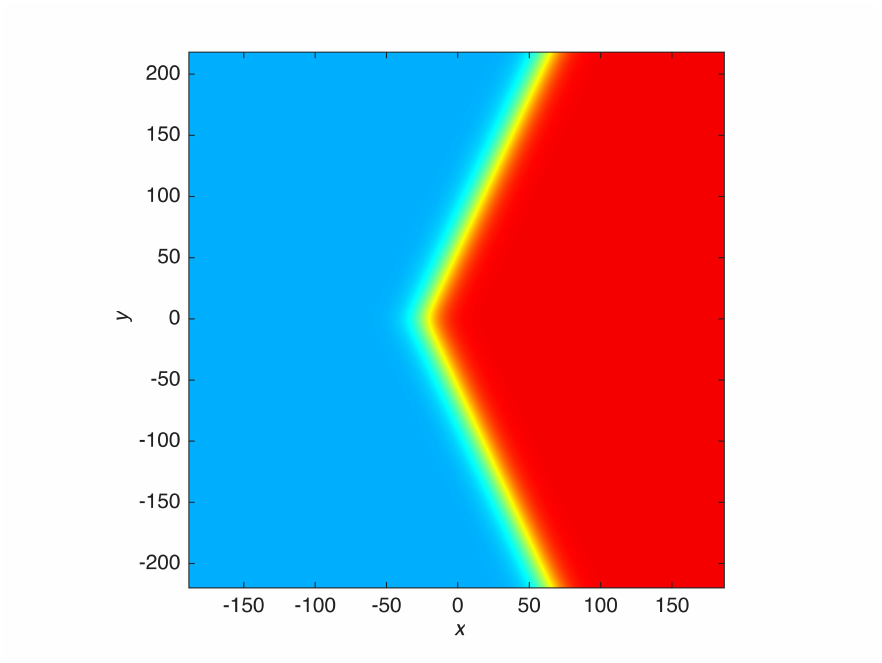}
\includegraphics[height=0.25\textwidth,trim=40 0 40 0,clip]{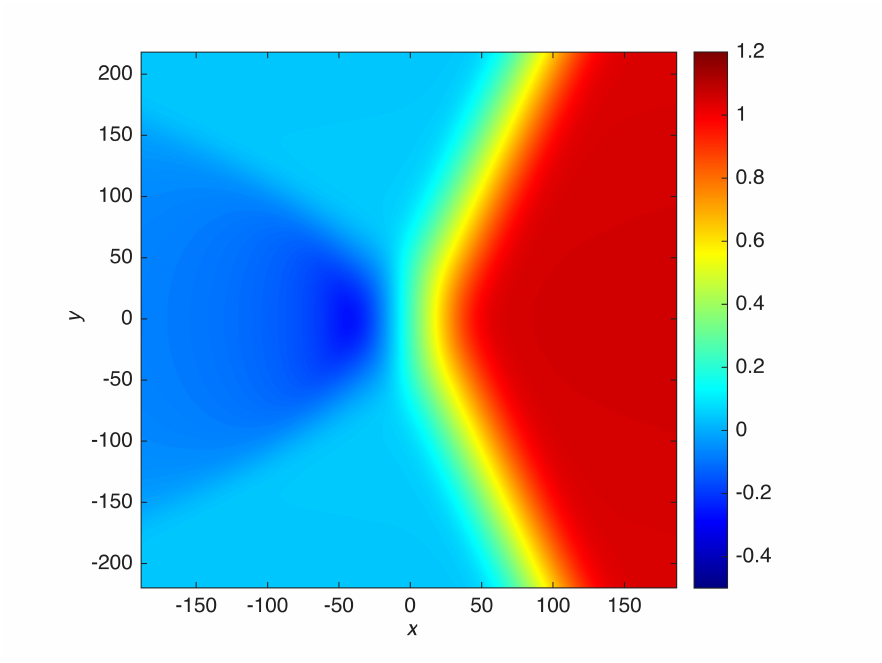}}
\kern-\medskipamount
\caption{Same as Fig.~\ref{f:W3sub}, but for a type IV wedge with $q_o = - 0.4$.}
\label{f:W4sub}
\vskip\medskipamount
\centerline{\includegraphics[height=0.25\textwidth,trim=40 0 40 0,clip]{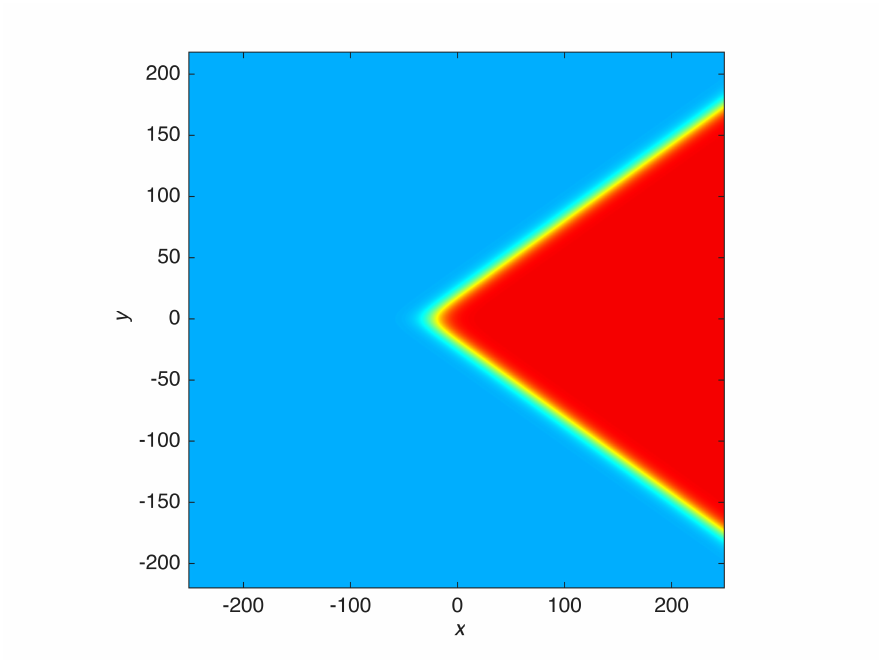}
\includegraphics[height=0.25\textwidth,trim=40 0 40 0,clip]{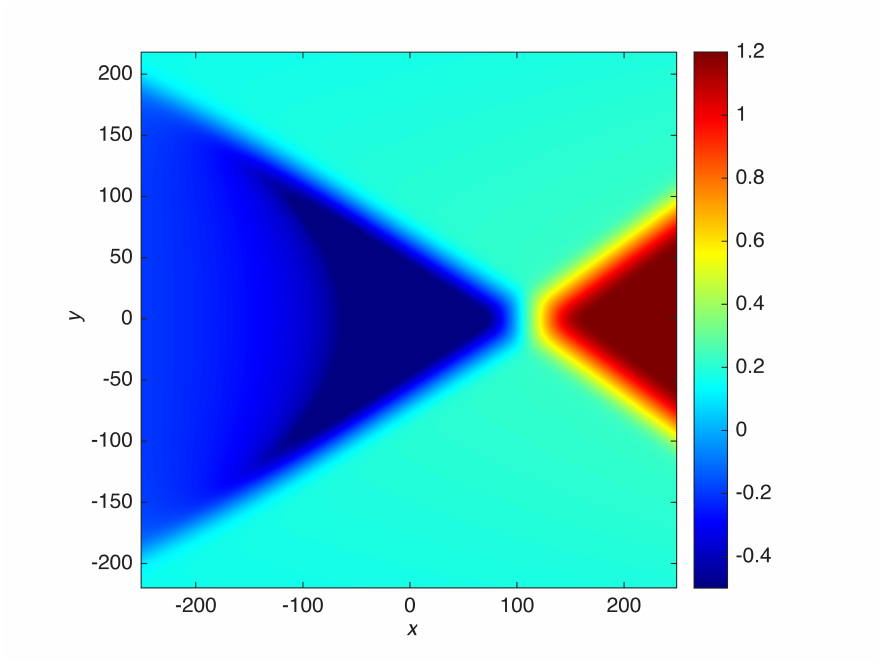}}
\kern-\medskipamount
\caption{Same as Fig~.\ref{f:W4sub} but for $q_o = - 1.6$.}
\label{f:W4super}
\kern-\medskipamount
\end{figure}

Given these large domains, the number of Fourier modes (equivalently, the number of spatial grid points) 
must be taken to be correspondingly large in order to maintain spatial accuracy. 
We choose the number of Fourier modes along the $x$ direction and the $y$ direction to be respectively $2^{15}$ and $2^{13}$, 
which results in a spatial domain with grid spacings $\Delta x=\Delta y=0.3052$ and a corresponding Fourier domain with grid spacings 
$\Delta k=0.00062832$ and $\Delta l=0.0025$. 
We choose a corresponding time integration step size of $\Delta t=10^{-2}$ 
to ensure stable and accurate results. 
The deviation of the solution at $x = \pm x_\mathrm{max}$ from the unperturbed zero value is monitored at all times in each case to verify the accuracy of the results.

\textbf{Implementation.}
The simulations presented in this work were carried out in Matlab on a cluster computing network
at the Center for Computational Research (CCR) at the University at Buffalo.

The large number of Fourier modes, coupled with the multiple FFTs required at each time step, 
made it necessary to use GPU acceleration techniques. 
(While a computer's CPU typically has more powerful processing cores, it only has a few of them. 
Conversely, a GPU has tens or hundreds of processing cores, which makes it an excellent tool for parallel computing.) 
Care was taken to manage the amount of memory available on the GPUs
(since exceeding the allotted memory of the GPU would result in an early termination of the simulation) 
as well as to minimize the transfers of information between the GPU and the CPU 
(since the transfer of data between CPU and GPU is a slow process on current hardware, which 
results in a bottleneck if too many transfers are made). 
Accordingly, once the parameters for the simulation have been used to create the initial conditions for $\hat u$ and $\hat v$, all parameters necessary to evolve $\hat v$ are transferred to the GPU. 
Then we take the IFFT of $\hat u$ and transfer it to the CPU to be written to disk only when necessary to save each temporal snapshot.

Given the kind of GPUs available, the vast majority of the simulation time is due to data transfer between the GPU and the CPU as well as time spent saving data to disk. 
A typical simulation with the parameter values described above required about an hour to be completed. 

\textbf{Dynamics of supercritical type I wedges.}
Figure~\ref{f:W1super} shows the temporal evolution of a supercritical type~I wedge.
This should be compared to Fig.~\ref{f:W1sub} in the main text, which is for a subcritical type~I wedge.
Although the dynamics in the two cases is qualitatively similar,
important quantitative difference exist.
Namely, in the supercritical case the solution approaches a curved DSW of cylindrical KdV, and 
amplitude of the oscillations near $y=0$ correspondingly tends to zero as $t\to\infty$,
as discussed in the main text.

\textbf{Dynamics of type III and type IV wedges.}
The temporal evolution of wedge types III and IV does not result in the generation
of DSWs, and gives rise instead to two-dimensional rarefaction waves,
as shown in Figs.~\ref{f:W3sub}--\ref{f:W4super}.
Two interesting questions in this respect are 
(i) whether or not there are qualitative differences between subcritical and supercritical type~III and type~IV wedges (similarly to type I and~II wedges),
and 
(ii) whether it is possible to quantitatively characterize this behavior via suitable solutions
of the dispersionless KP equation.
These questions are left for a future investigation.

\vfill

%% file: kpwedge_arxiv.bbl
\begin{thebibliography}{81}%
\makeatletter
\providecommand \@ifxundefined [1]{%
 \@ifx{#1\undefined}
}%
\providecommand \@ifnum [1]{%
 \ifnum #1\expandafter \@firstoftwo
 \else \expandafter \@secondoftwo
 \fi
}%
\providecommand \@ifx [1]{%
 \ifx #1\expandafter \@firstoftwo
 \else \expandafter \@secondoftwo
 \fi
}%
\providecommand \natexlab [1]{#1}%
\providecommand \enquote  [1]{``#1''}%
\providecommand \bibnamefont  [1]{#1}%
\providecommand \bibfnamefont [1]{#1}%
\providecommand \citenamefont [1]{#1}%
\providecommand \href@noop [0]{\@secondoftwo}%
\providecommand \href [0]{\begingroup \@sanitize@url \@href}%
\providecommand \@href[1]{\@@startlink{#1}\@@href}%
\providecommand \@@href[1]{\endgroup#1\@@endlink}%
\providecommand \@sanitize@url [0]{\catcode `\\12\catcode `\$12\catcode
  `\&12\catcode `\#12\catcode `\^12\catcode `\_12\catcode `\%12\relax}%
\providecommand \@@startlink[1]{}%
\providecommand \@@endlink[0]{}%
\providecommand \url  [0]{\begingroup\@sanitize@url \@url }%
\providecommand \@url [1]{\endgroup\@href {#1}{\urlprefix }}%
\providecommand \urlprefix  [0]{URL }%
\providecommand \Eprint [0]{\href }%
\providecommand \doibase [0]{http://dx.doi.org/}%
\providecommand \selectlanguage [0]{\@gobble}%
\providecommand \bibinfo  [0]{\@secondoftwo}%
\providecommand \bibfield  [0]{\@secondoftwo}%
\providecommand \translation [1]{[#1]}%
\providecommand \BibitemOpen [0]{}%
\providecommand \bibitemStop [0]{}%
\providecommand \bibitemNoStop [0]{.\EOS\space}%
\providecommand \EOS [0]{\spacefactor3000\relax}%
\providecommand \BibitemShut  [1]{\csname bibitem#1\endcsname}%
\let\auto@bib@innerbib\@empty
\bibitem [{\citenamefont {Wan}\ \emph {et~al.}(2007)\citenamefont {Wan},
  \citenamefont {Jia},\ and\ \citenamefont {Fleischer}}]{wan_dispersive_2007}%
  \BibitemOpen
  \bibfield  {author} {\bibinfo {author} {\bibfnamefont {W.}~\bibnamefont
  {Wan}}, \bibinfo {author} {\bibfnamefont {S.}~\bibnamefont {Jia}}, \ and\
  \bibinfo {author} {\bibfnamefont {J.~W.}\ \bibnamefont {Fleischer}},\
  }\href@noop {} {\bibfield  {journal} {\bibinfo  {journal} {Nat Phys}\
  }\textbf {\bibinfo {volume} {3}},\ \bibinfo {pages} {46} (\bibinfo {year}
  {2007})}\BibitemShut {NoStop}%
\bibitem [{\citenamefont {Xu}\ \emph {et~al.}(2017)\citenamefont {Xu},
  \citenamefont {Conforti}, \citenamefont {Kudlinski}, \citenamefont {Mussot},\
  and\ \citenamefont {Trillo}}]{xu_dispersive_2017}%
  \BibitemOpen
  \bibfield  {author} {\bibinfo {author} {\bibfnamefont {G.}~\bibnamefont
  {Xu}}, \bibinfo {author} {\bibfnamefont {M.}~\bibnamefont {Conforti}},
  \bibinfo {author} {\bibfnamefont {A.}~\bibnamefont {Kudlinski}}, \bibinfo
  {author} {\bibfnamefont {A.}~\bibnamefont {Mussot}}, \ and\ \bibinfo {author}
  {\bibfnamefont {S.}~\bibnamefont {Trillo}},\ }\href@noop {} {\bibfield
  {journal} {\bibinfo  {journal} {Phys. Rev. Lett.}\ }\textbf {\bibinfo
  {volume} {118}},\ \bibinfo {pages} {254101} (\bibinfo {year}
  {2017})}\BibitemShut {NoStop}%
\bibitem [{\citenamefont {Bienaim{\'e}}\ \emph {et~al.}(2021)\citenamefont
  {Bienaim{\'e}}, \citenamefont {Isoard}, \citenamefont {Fontaine},
  \citenamefont {Bramati}, \citenamefont {Kamchatnov}, \citenamefont
  {Glorieux},\ and\ \citenamefont {Pavloff}}]{bienaime_quantitative_2021}%
  \BibitemOpen
  \bibfield  {author} {\bibinfo {author} {\bibfnamefont {T.}~\bibnamefont
  {Bienaim{\'e}}}, \bibinfo {author} {\bibfnamefont {M.}~\bibnamefont
  {Isoard}}, \bibinfo {author} {\bibfnamefont {Q.}~\bibnamefont {Fontaine}},
  \bibinfo {author} {\bibfnamefont {A.}~\bibnamefont {Bramati}}, \bibinfo
  {author} {\bibfnamefont {A.~M.}\ \bibnamefont {Kamchatnov}}, \bibinfo
  {author} {\bibfnamefont {Q.}~\bibnamefont {Glorieux}}, \ and\ \bibinfo
  {author} {\bibfnamefont {N.}~\bibnamefont {Pavloff}},\ }\href@noop {}
  {\bibfield  {journal} {\bibinfo  {journal} {Phys. Rev. Lett.}\ }\textbf
  {\bibinfo {volume} {126}},\ \bibinfo {pages} {183901} (\bibinfo {year}
  {2021})}\BibitemShut {NoStop}%
\bibitem [{\citenamefont {Bendahmane}\ \emph {et~al.}(2022)\citenamefont
  {Bendahmane}, \citenamefont {Xu}, \citenamefont {Conforti}, \citenamefont
  {Kudlinski}, \citenamefont {Mussot},\ and\ \citenamefont
  {Trillo}}]{bendahmane_piston_2022}%
  \BibitemOpen
  \bibfield  {author} {\bibinfo {author} {\bibfnamefont {A.}~\bibnamefont
  {Bendahmane}}, \bibinfo {author} {\bibfnamefont {G.}~\bibnamefont {Xu}},
  \bibinfo {author} {\bibfnamefont {M.}~\bibnamefont {Conforti}}, \bibinfo
  {author} {\bibfnamefont {A.}~\bibnamefont {Kudlinski}}, \bibinfo {author}
  {\bibfnamefont {A.}~\bibnamefont {Mussot}}, \ and\ \bibinfo {author}
  {\bibfnamefont {S.}~\bibnamefont {Trillo}},\ }\href@noop {} {\bibfield
  {journal} {\bibinfo  {journal} {Nat Commun}\ }\textbf {\bibinfo {volume}
  {13}},\ \bibinfo {pages} {3137} (\bibinfo {year} {2022})}\BibitemShut
  {NoStop}%
\bibitem [{\citenamefont {Dutton}\ \emph {et~al.}(2001)\citenamefont {Dutton},
  \citenamefont {Budde}, \citenamefont {Slowe},\ and\ \citenamefont
  {Hau}}]{dutton_observation_2001}%
  \BibitemOpen
  \bibfield  {author} {\bibinfo {author} {\bibfnamefont {Z.}~\bibnamefont
  {Dutton}}, \bibinfo {author} {\bibfnamefont {M.}~\bibnamefont {Budde}},
  \bibinfo {author} {\bibfnamefont {C.}~\bibnamefont {Slowe}}, \ and\ \bibinfo
  {author} {\bibfnamefont {L.~V.}\ \bibnamefont {Hau}},\ }\href@noop {}
  {\bibfield  {journal} {\bibinfo  {journal} {Science}\ }\textbf {\bibinfo
  {volume} {293}},\ \bibinfo {pages} {663} (\bibinfo {year}
  {2001})}\BibitemShut {NoStop}%
\bibitem [{\citenamefont {Hoefer}\ \emph {et~al.}(2006)\citenamefont {Hoefer},
  \citenamefont {Ablowitz}, \citenamefont {Coddington}, \citenamefont
  {Cornell}, \citenamefont {Engels},\ and\ \citenamefont
  {Schweikhard}}]{hoefer_dispersive_2006}%
  \BibitemOpen
  \bibfield  {author} {\bibinfo {author} {\bibfnamefont {M.~A.}\ \bibnamefont
  {Hoefer}}, \bibinfo {author} {\bibfnamefont {M.~J.}\ \bibnamefont
  {Ablowitz}}, \bibinfo {author} {\bibfnamefont {I.}~\bibnamefont
  {Coddington}}, \bibinfo {author} {\bibfnamefont {E.~A.}\ \bibnamefont
  {Cornell}}, \bibinfo {author} {\bibfnamefont {P.}~\bibnamefont {Engels}}, \
  and\ \bibinfo {author} {\bibfnamefont {V.}~\bibnamefont {Schweikhard}},\
  }\href@noop {} {\bibfield  {journal} {\bibinfo  {journal} {Phys. Rev. A}\
  }\textbf {\bibinfo {volume} {74}},\ \bibinfo {pages} {023623} (\bibinfo
  {year} {2006})}\BibitemShut {NoStop}%
\bibitem [{\citenamefont {Joseph}\ \emph {et~al.}(2011)\citenamefont {Joseph},
  \citenamefont {Thomas}, \citenamefont {Kulkarni},\ and\ \citenamefont
  {Abanov}}]{joseph_observation_2011}%
  \BibitemOpen
  \bibfield  {author} {\bibinfo {author} {\bibfnamefont {J.~A.}\ \bibnamefont
  {Joseph}}, \bibinfo {author} {\bibfnamefont {J.~E.}\ \bibnamefont {Thomas}},
  \bibinfo {author} {\bibfnamefont {M.}~\bibnamefont {Kulkarni}}, \ and\
  \bibinfo {author} {\bibfnamefont {A.~G.}\ \bibnamefont {Abanov}},\
  }\href@noop {} {\bibfield  {journal} {\bibinfo  {journal} {Phys. Rev. Lett.}\
  }\textbf {\bibinfo {volume} {106}},\ \bibinfo {pages} {150401} (\bibinfo
  {year} {2011})}\BibitemShut {NoStop}%
\bibitem [{\citenamefont {Mossman}\ \emph {et~al.}(2018)\citenamefont
  {Mossman}, \citenamefont {Hoefer}, \citenamefont {Julien}, \citenamefont
  {Kevrekidis},\ and\ \citenamefont {Engels}}]{mossman_dissipative_2018}%
  \BibitemOpen
  \bibfield  {author} {\bibinfo {author} {\bibfnamefont {M.~E.}\ \bibnamefont
  {Mossman}}, \bibinfo {author} {\bibfnamefont {M.~A.}\ \bibnamefont {Hoefer}},
  \bibinfo {author} {\bibfnamefont {K.}~\bibnamefont {Julien}}, \bibinfo
  {author} {\bibfnamefont {P.~G.}\ \bibnamefont {Kevrekidis}}, \ and\ \bibinfo
  {author} {\bibfnamefont {P.}~\bibnamefont {Engels}},\ }\href@noop {}
  {\bibfield  {journal} {\bibinfo  {journal} {Nat. Commun.}\ }\textbf {\bibinfo
  {volume} {9}},\ \bibinfo {pages} {4665} (\bibinfo {year} {2018})}\BibitemShut
  {NoStop}%
\bibitem [{\citenamefont {Nash}\ and\ \citenamefont
  {Moum}(2005)}]{nash_river_2005}%
  \BibitemOpen
  \bibfield  {author} {\bibinfo {author} {\bibfnamefont {J.~D.}\ \bibnamefont
  {Nash}}\ and\ \bibinfo {author} {\bibfnamefont {J.~N.}\ \bibnamefont
  {Moum}},\ }\href@noop {} {\bibfield  {journal} {\bibinfo  {journal} {Nature}\
  }\textbf {\bibinfo {volume} {437}},\ \bibinfo {pages} {400} (\bibinfo {year}
  {2005})}\BibitemShut {NoStop}%
\bibitem [{\citenamefont {Scotti}\ \emph {et~al.}(2007)\citenamefont {Scotti},
  \citenamefont {Beardsley},\ and\ \citenamefont
  {Butman}}]{scotti_generation_2007}%
  \BibitemOpen
  \bibfield  {author} {\bibinfo {author} {\bibfnamefont {A.}~\bibnamefont
  {Scotti}}, \bibinfo {author} {\bibfnamefont {R.~C.}\ \bibnamefont
  {Beardsley}}, \ and\ \bibinfo {author} {\bibfnamefont {B.}~\bibnamefont
  {Butman}},\ }\href@noop {} {\bibfield  {journal} {\bibinfo  {journal} {J.
  Geophys. Res.}\ }\textbf {\bibinfo {volume} {112}} (\bibinfo {year}
  {2007})}\BibitemShut {NoStop}%
\bibitem [{\citenamefont {Trillo}\ \emph
  {et~al.}(2016{\natexlab{a}})\citenamefont {Trillo}, \citenamefont {Deng},
  \citenamefont {Biondini}, \citenamefont {Klein}, \citenamefont {Clauss},
  \citenamefont {Chabchoub},\ and\ \citenamefont
  {Onorato}}]{trillo_experimental_2016}%
  \BibitemOpen
  \bibfield  {author} {\bibinfo {author} {\bibfnamefont {S.}~\bibnamefont
  {Trillo}}, \bibinfo {author} {\bibfnamefont {G.}~\bibnamefont {Deng}},
  \bibinfo {author} {\bibfnamefont {G.}~\bibnamefont {Biondini}}, \bibinfo
  {author} {\bibfnamefont {M.}~\bibnamefont {Klein}}, \bibinfo {author}
  {\bibfnamefont {G.~F.}\ \bibnamefont {Clauss}}, \bibinfo {author}
  {\bibfnamefont {A.}~\bibnamefont {Chabchoub}}, \ and\ \bibinfo {author}
  {\bibfnamefont {M.}~\bibnamefont {Onorato}},\ }\href@noop {} {\bibfield
  {journal} {\bibinfo  {journal} {Phys. Rev. Lett.}\ }\textbf {\bibinfo
  {volume} {117}},\ \bibinfo {pages} {144102} (\bibinfo {year}
  {2016}{\natexlab{a}})}\BibitemShut {NoStop}%
\bibitem [{\citenamefont {Li}\ \emph {et~al.}(2018)\citenamefont {Li},
  \citenamefont {Pawlowicz},\ and\ \citenamefont {Wang}}]{li_seasonal_2018}%
  \BibitemOpen
  \bibfield  {author} {\bibinfo {author} {\bibfnamefont {L.}~\bibnamefont
  {Li}}, \bibinfo {author} {\bibfnamefont {R.}~\bibnamefont {Pawlowicz}}, \
  and\ \bibinfo {author} {\bibfnamefont {C.}~\bibnamefont {Wang}},\ }\href@noop
  {} {\bibfield  {journal} {\bibinfo  {journal} {Journal of Geophysical
  Research: Oceans}\ }\textbf {\bibinfo {volume} {123}},\ \bibinfo {pages}
  {5706} (\bibinfo {year} {2018})}\BibitemShut {NoStop}%
\bibitem [{\citenamefont {Mo}\ \emph {et~al.}(2013)\citenamefont {Mo},
  \citenamefont {Kishek}, \citenamefont {Feldman}, \citenamefont {Haber},
  \citenamefont {Beaudoin}, \citenamefont {O'Shea},\ and\ \citenamefont
  {Thangaraj}}]{mo_experimental_2013}%
  \BibitemOpen
  \bibfield  {author} {\bibinfo {author} {\bibfnamefont {Y.~C.}\ \bibnamefont
  {Mo}}, \bibinfo {author} {\bibfnamefont {R.~A.}\ \bibnamefont {Kishek}},
  \bibinfo {author} {\bibfnamefont {D.}~\bibnamefont {Feldman}}, \bibinfo
  {author} {\bibfnamefont {I.}~\bibnamefont {Haber}}, \bibinfo {author}
  {\bibfnamefont {B.}~\bibnamefont {Beaudoin}}, \bibinfo {author}
  {\bibfnamefont {P.~G.}\ \bibnamefont {O'Shea}}, \ and\ \bibinfo {author}
  {\bibfnamefont {J.~C.~T.}\ \bibnamefont {Thangaraj}},\ }\href@noop {}
  {\bibfield  {journal} {\bibinfo  {journal} {Phys. Rev. Lett.}\ }\textbf
  {\bibinfo {volume} {110}},\ \bibinfo {pages} {084802} (\bibinfo {year}
  {2013})}\BibitemShut {NoStop}%
\bibitem [{\citenamefont {Maiden}\ \emph {et~al.}(2016)\citenamefont {Maiden},
  \citenamefont {Lowman}, \citenamefont {Anderson}, \citenamefont {Schubert},\
  and\ \citenamefont {Hoefer}}]{maiden_observation_2016}%
  \BibitemOpen
  \bibfield  {author} {\bibinfo {author} {\bibfnamefont {M.~D.}\ \bibnamefont
  {Maiden}}, \bibinfo {author} {\bibfnamefont {N.~K.}\ \bibnamefont {Lowman}},
  \bibinfo {author} {\bibfnamefont {D.~V.}\ \bibnamefont {Anderson}}, \bibinfo
  {author} {\bibfnamefont {M.~E.}\ \bibnamefont {Schubert}}, \ and\ \bibinfo
  {author} {\bibfnamefont {M.~A.}\ \bibnamefont {Hoefer}},\ }\href@noop {}
  {\bibfield  {journal} {\bibinfo  {journal} {Phys. Rev. Lett.}\ }\textbf
  {\bibinfo {volume} {116}},\ \bibinfo {pages} {174501} (\bibinfo {year}
  {2016})}\BibitemShut {NoStop}%
\bibitem [{\citenamefont {Janantha}\ \emph {et~al.}(2017)\citenamefont
  {Janantha}, \citenamefont {Sprenger}, \citenamefont {Hoefer},\ and\
  \citenamefont {Wu}}]{janantha_observation_2017}%
  \BibitemOpen
  \bibfield  {author} {\bibinfo {author} {\bibfnamefont {P.~A.~P.}\
  \bibnamefont {Janantha}}, \bibinfo {author} {\bibfnamefont {P.}~\bibnamefont
  {Sprenger}}, \bibinfo {author} {\bibfnamefont {M.~A.}\ \bibnamefont
  {Hoefer}}, \ and\ \bibinfo {author} {\bibfnamefont {M.}~\bibnamefont {Wu}},\
  }\href@noop {} {\bibfield  {journal} {\bibinfo  {journal} {Phys. Rev. Lett.}\
  }\textbf {\bibinfo {volume} {119}},\ \bibinfo {pages} {024101} (\bibinfo
  {year} {2017})}\BibitemShut {NoStop}%
\bibitem [{\citenamefont {Li}\ \emph {et~al.}(2021)\citenamefont {Li},
  \citenamefont {Chockalingam},\ and\ \citenamefont
  {Cohen}}]{li_observation_2021}%
  \BibitemOpen
  \bibfield  {author} {\bibinfo {author} {\bibfnamefont {J.}~\bibnamefont
  {Li}}, \bibinfo {author} {\bibfnamefont {S.}~\bibnamefont {Chockalingam}}, \
  and\ \bibinfo {author} {\bibfnamefont {T.}~\bibnamefont {Cohen}},\
  }\href@noop {} {\bibfield  {journal} {\bibinfo  {journal} {Phys. Rev. Lett.}\
  }\textbf {\bibinfo {volume} {127}},\ \bibinfo {pages} {014302} (\bibinfo
  {year} {2021})}\BibitemShut {NoStop}%
\bibitem [{\citenamefont {Trillo}\ \emph
  {et~al.}(2016{\natexlab{b}})\citenamefont {Trillo}, \citenamefont {Klein},
  \citenamefont {Clauss},\ and\ \citenamefont
  {Onorato}}]{trillo_observation_2016}%
  \BibitemOpen
  \bibfield  {author} {\bibinfo {author} {\bibfnamefont {S.}~\bibnamefont
  {Trillo}}, \bibinfo {author} {\bibfnamefont {M.}~\bibnamefont {Klein}},
  \bibinfo {author} {\bibfnamefont {G.}~\bibnamefont {Clauss}}, \ and\ \bibinfo
  {author} {\bibfnamefont {M.}~\bibnamefont {Onorato}},\ }\href@noop {}
  {\bibfield  {journal} {\bibinfo  {journal} {Phys. Nonlinear Phenom.}\
  }\textbf {\bibinfo {volume} {333}},\ \bibinfo {pages} {276} (\bibinfo {year}
  {2016}{\natexlab{b}})}\BibitemShut {NoStop}%
\bibitem [{\citenamefont {Chassagne}\ \emph {et~al.}(2019)\citenamefont
  {Chassagne}, \citenamefont {Filippini}, \citenamefont {Ricchiuto},\ and\
  \citenamefont {Bonneton}}]{chassagne_dispersive_2019}%
  \BibitemOpen
  \bibfield  {author} {\bibinfo {author} {\bibfnamefont {R.}~\bibnamefont
  {Chassagne}}, \bibinfo {author} {\bibfnamefont {A.~G.}\ \bibnamefont
  {Filippini}}, \bibinfo {author} {\bibfnamefont {M.}~\bibnamefont
  {Ricchiuto}}, \ and\ \bibinfo {author} {\bibfnamefont {P.}~\bibnamefont
  {Bonneton}},\ }\href@noop {} {\bibfield  {journal} {\bibinfo  {journal} {J.
  Fluid Mech.}\ }\textbf {\bibinfo {volume} {870}},\ \bibinfo {pages} {595}
  (\bibinfo {year} {2019})}\BibitemShut {NoStop}%
\bibitem [{\citenamefont {Simmons}\ \emph {et~al.}(2020)\citenamefont
  {Simmons}, \citenamefont {Bayocboc}, \citenamefont {Pillay}, \citenamefont
  {Colas}, \citenamefont {McCulloch},\ and\ \citenamefont
  {Kheruntsyan}}]{simmons_what_2020}%
  \BibitemOpen
  \bibfield  {author} {\bibinfo {author} {\bibfnamefont {S.~A.}\ \bibnamefont
  {Simmons}}, \bibinfo {author} {\bibfnamefont {F.~A.}\ \bibnamefont
  {Bayocboc}}, \bibinfo {author} {\bibfnamefont {J.~C.}\ \bibnamefont
  {Pillay}}, \bibinfo {author} {\bibfnamefont {D.}~\bibnamefont {Colas}},
  \bibinfo {author} {\bibfnamefont {I.~P.}\ \bibnamefont {McCulloch}}, \ and\
  \bibinfo {author} {\bibfnamefont {K.~V.}\ \bibnamefont {Kheruntsyan}},\
  }\href@noop {} {\bibfield  {journal} {\bibinfo  {journal} {Phys. Rev. Lett.}\
  }\textbf {\bibinfo {volume} {125}},\ \bibinfo {pages} {180401} (\bibinfo
  {year} {2020})}\BibitemShut {NoStop}%
\bibitem [{\citenamefont {El}\ and\ \citenamefont
  {Hoefer}(2016)}]{PHYSD333p11}%
  \BibitemOpen
  \bibfield  {author} {\bibinfo {author} {\bibfnamefont {G.}~\bibnamefont
  {El}}\ and\ \bibinfo {author} {\bibfnamefont {M.~A.}\ \bibnamefont
  {Hoefer}},\ }\href@noop {} {\bibfield  {journal} {\bibinfo  {journal} {Phys.
  D}\ }\textbf {\bibinfo {volume} {333}},\ \bibinfo {pages} {11} (\bibinfo
  {year} {2016})}\BibitemShut {NoStop}%
\bibitem [{\citenamefont {Miller}(2016)}]{miller_generation_2016}%
  \BibitemOpen
  \bibfield  {author} {\bibinfo {author} {\bibfnamefont {P.~D.}\ \bibnamefont
  {Miller}},\ }\href@noop {} {\bibfield  {journal} {\bibinfo  {journal} {Phys.
  Nonlinear Phenom.}\ }\textbf {\bibinfo {volume} {333}},\ \bibinfo {pages}
  {66} (\bibinfo {year} {2016})}\BibitemShut {NoStop}%
\bibitem [{\citenamefont {Ablowitz}\ \emph {et~al.}(2016)\citenamefont
  {Ablowitz}, \citenamefont {Demirci},\ and\ \citenamefont
  {Ma}}]{ablowitz_dispersive_2016}%
  \BibitemOpen
  \bibfield  {author} {\bibinfo {author} {\bibfnamefont {M.~J.}\ \bibnamefont
  {Ablowitz}}, \bibinfo {author} {\bibfnamefont {A.}~\bibnamefont {Demirci}}, \
  and\ \bibinfo {author} {\bibfnamefont {Y.-P.}\ \bibnamefont {Ma}},\
  }\href@noop {} {\bibfield  {journal} {\bibinfo  {journal} {Phys. Nonlinear
  Phenom.}\ }\textbf {\bibinfo {volume} {333}},\ \bibinfo {pages} {84}
  (\bibinfo {year} {2016})}\BibitemShut {NoStop}%
\bibitem [{\citenamefont {Gurevich}\ \emph {et~al.}(1995)\citenamefont
  {Gurevich}, \citenamefont {Krylov}, \citenamefont {Khodorovskii},\ and\
  \citenamefont {El}}]{gurevich_supersonic_1995}%
  \BibitemOpen
  \bibfield  {author} {\bibinfo {author} {\bibfnamefont {A.~V.}\ \bibnamefont
  {Gurevich}}, \bibinfo {author} {\bibfnamefont {A.~L.}\ \bibnamefont
  {Krylov}}, \bibinfo {author} {\bibfnamefont {V.~V.}\ \bibnamefont
  {Khodorovskii}}, \ and\ \bibinfo {author} {\bibfnamefont {G.~A.}\
  \bibnamefont {El}},\ }\href@noop {} {\bibfield  {journal} {\bibinfo
  {journal} {Sov. Phys. JETP}\ }\textbf {\bibinfo {volume} {81}},\ \bibinfo
  {pages} {87} (\bibinfo {year} {1995})}\BibitemShut {NoStop}%
\bibitem [{\citenamefont {Gurevich}\ \emph {et~al.}(1996)\citenamefont
  {Gurevich}, \citenamefont {Krylov}, \citenamefont {Khodorovskii},\ and\
  \citenamefont {El}}]{gurevich_supersonic_1996}%
  \BibitemOpen
  \bibfield  {author} {\bibinfo {author} {\bibfnamefont {A.~V.}\ \bibnamefont
  {Gurevich}}, \bibinfo {author} {\bibfnamefont {A.~L.}\ \bibnamefont
  {Krylov}}, \bibinfo {author} {\bibfnamefont {V.~V.}\ \bibnamefont
  {Khodorovskii}}, \ and\ \bibinfo {author} {\bibfnamefont {G.~A.}\
  \bibnamefont {El}},\ }\href@noop {} {\bibfield  {journal} {\bibinfo
  {journal} {Sov. Phys. JETP}\ }\textbf {\bibinfo {volume} {82}},\ \bibinfo
  {pages} {709} (\bibinfo {year} {1996})}\BibitemShut {NoStop}%
\bibitem [{\citenamefont {El}\ \emph {et~al.}(2009)\citenamefont {El},
  \citenamefont {Kamchatnov}, \citenamefont {Khodorovskii}, \citenamefont
  {Annibale},\ and\ \citenamefont {Gammal}}]{el_two-dimensional_2009}%
  \BibitemOpen
  \bibfield  {author} {\bibinfo {author} {\bibfnamefont {G.~A.}\ \bibnamefont
  {El}}, \bibinfo {author} {\bibfnamefont {A.~M.}\ \bibnamefont {Kamchatnov}},
  \bibinfo {author} {\bibfnamefont {V.~V.}\ \bibnamefont {Khodorovskii}},
  \bibinfo {author} {\bibfnamefont {E.~S.}\ \bibnamefont {Annibale}}, \ and\
  \bibinfo {author} {\bibfnamefont {A.}~\bibnamefont {Gammal}},\ }\href@noop {}
  {\bibfield  {journal} {\bibinfo  {journal} {Phys. Rev. E}\ }\textbf {\bibinfo
  {volume} {80}},\ \bibinfo {pages} {046317} (\bibinfo {year}
  {2009})}\BibitemShut {NoStop}%
\bibitem [{\citenamefont {Hoefer}\ and\ \citenamefont
  {Ilan}(2012)}]{hoefer_dark_2012}%
  \BibitemOpen
  \bibfield  {author} {\bibinfo {author} {\bibfnamefont {M.~A.}\ \bibnamefont
  {Hoefer}}\ and\ \bibinfo {author} {\bibfnamefont {B.}~\bibnamefont {Ilan}},\
  }\href@noop {} {\bibfield  {journal} {\bibinfo  {journal} {SIAM Multiscale
  Model. Simul.}\ }\textbf {\bibinfo {volume} {10}},\ \bibinfo {pages} {306}
  (\bibinfo {year} {2012})}\BibitemShut {NoStop}%
\bibitem [{\citenamefont {Hoefer}\ \emph {et~al.}(2017)\citenamefont {Hoefer},
  \citenamefont {El},\ and\ \citenamefont {Kamchatnov}}]{hoefer_oblique_2017}%
  \BibitemOpen
  \bibfield  {author} {\bibinfo {author} {\bibfnamefont {M.~A.}\ \bibnamefont
  {Hoefer}}, \bibinfo {author} {\bibfnamefont {G.~A.}\ \bibnamefont {El}}, \
  and\ \bibinfo {author} {\bibfnamefont {A.~M.}\ \bibnamefont {Kamchatnov}},\
  }\href@noop {} {\bibfield  {journal} {\bibinfo  {journal} {SIAM J. Appl.
  Math.}\ }\textbf {\bibinfo {volume} {77}},\ \bibinfo {pages} {1352} (\bibinfo
  {year} {2017})}\BibitemShut {NoStop}%
\bibitem [{\citenamefont {Lax}(1973)}]{Lax1973}%
  \BibitemOpen
  \bibfield  {author} {\bibinfo {author} {\bibfnamefont {P.~D.}\ \bibnamefont
  {Lax}},\ }\href@noop {} {\emph {\bibinfo {title} {Hyperbolic systems of
  conservation laws and the mathematical theory of shock waves}}}\ (\bibinfo
  {publisher} {SIAM},\ \bibinfo {year} {1973})\BibitemShut {NoStop}%
\bibitem [{\citenamefont {Courant}\ and\ \citenamefont
  {Friedrichs}(1976)}]{CF1976}%
  \BibitemOpen
  \bibfield  {author} {\bibinfo {author} {\bibfnamefont {R.}~\bibnamefont
  {Courant}}\ and\ \bibinfo {author} {\bibfnamefont {K.~O.}\ \bibnamefont
  {Friedrichs}},\ }\href@noop {} {\emph {\bibinfo {title} {Supersonic flow and
  shock waves}}}\ (\bibinfo  {publisher} {Springer},\ \bibinfo {year}
  {1976})\BibitemShut {NoStop}%
\bibitem [{\citenamefont {Courant}\ and\ \citenamefont
  {Hilbert}(1962)}]{CH1962}%
  \BibitemOpen
  \bibfield  {author} {\bibinfo {author} {\bibfnamefont {R.}~\bibnamefont
  {Courant}}\ and\ \bibinfo {author} {\bibfnamefont {H.}~\bibnamefont
  {Hilbert}},\ }\href@noop {} {\emph {\bibinfo {title} {Methods of mathematical
  physics}}}\ (\bibinfo  {publisher} {Interscience},\ \bibinfo {address} {New
  York},\ \bibinfo {year} {1962})\BibitemShut {NoStop}%
\bibitem [{\citenamefont {Zhang}\ and\ \citenamefont
  {Zheng}(1990)}]{zhang_conjecture_1990}%
  \BibitemOpen
  \bibfield  {author} {\bibinfo {author} {\bibfnamefont {T.}~\bibnamefont
  {Zhang}}\ and\ \bibinfo {author} {\bibfnamefont {Y.~X.}\ \bibnamefont
  {Zheng}},\ }\href@noop {} {\bibfield  {journal} {\bibinfo  {journal} {SIAM J.
  Math. Anal.}\ }\textbf {\bibinfo {volume} {21}},\ \bibinfo {pages} {593}
  (\bibinfo {year} {1990})}\BibitemShut {NoStop}%
\bibitem [{\citenamefont {Kurganov}\ and\ \citenamefont
  {Tadmor}(2002)}]{kurganov_solution_2002}%
  \BibitemOpen
  \bibfield  {author} {\bibinfo {author} {\bibfnamefont {A.}~\bibnamefont
  {Kurganov}}\ and\ \bibinfo {author} {\bibfnamefont {E.}~\bibnamefont
  {Tadmor}},\ }\href@noop {} {\bibfield  {journal} {\bibinfo  {journal}
  {Numerical Methods Partial}\ }\textbf {\bibinfo {volume} {18}},\ \bibinfo
  {pages} {584} (\bibinfo {year} {2002})}\BibitemShut {NoStop}%
\bibitem [{\citenamefont {Miles}(1977)}]{JFM79p171}%
  \BibitemOpen
  \bibfield  {author} {\bibinfo {author} {\bibfnamefont {J.~W.}\ \bibnamefont
  {Miles}},\ }\href@noop {} {\bibfield  {journal} {\bibinfo  {journal} {J.
  Fluid Mech.}\ }\textbf {\bibinfo {volume} {79}},\ \bibinfo {pages} {171}
  (\bibinfo {year} {1977})}\BibitemShut {NoStop}%
\bibitem [{\citenamefont {Brio}\ and\ \citenamefont
  {Hunter}(1992)}]{brio_mach_1992}%
  \BibitemOpen
  \bibfield  {author} {\bibinfo {author} {\bibfnamefont {M.}~\bibnamefont
  {Brio}}\ and\ \bibinfo {author} {\bibfnamefont {J.~K.}\ \bibnamefont
  {Hunter}},\ }\href@noop {} {\bibfield  {journal} {\bibinfo  {journal} {Phys.
  D}\ }\textbf {\bibinfo {volume} {60}},\ \bibinfo {pages} {194} (\bibinfo
  {year} {1992})}\BibitemShut {NoStop}%
\bibitem [{\citenamefont {Scott~Russell}(1844)}]{Russell1844}%
  \BibitemOpen
  \bibfield  {author} {\bibinfo {author} {\bibfnamefont {J.}~\bibnamefont
  {Scott~Russell}},\ }\href@noop {} {\bibfield  {journal} {\bibinfo  {journal}
  {Rep. Meet. Brit. Assoc. Adv. Sci.}\ }\textbf {\bibinfo {volume} {14}},\
  \bibinfo {pages} {311} (\bibinfo {year} {1844})}\BibitemShut {NoStop}%
\bibitem [{\citenamefont {Kadomtsev}\ and\ \citenamefont
  {Petviashvili}(1970)}]{SovPhysDoklady15p539}%
  \BibitemOpen
  \bibfield  {author} {\bibinfo {author} {\bibfnamefont {B.~B.}\ \bibnamefont
  {Kadomtsev}}\ and\ \bibinfo {author} {\bibfnamefont {V.~I.}\ \bibnamefont
  {Petviashvili}},\ }\href@noop {} {\bibfield  {journal} {\bibinfo  {journal}
  {Sov. Phys. Doklady}\ }\textbf {\bibinfo {volume} {15}},\ \bibinfo {pages}
  {539} (\bibinfo {year} {1970})}\BibitemShut {NoStop}%
\bibitem [{\citenamefont {Ablowitz}\ and\ \citenamefont
  {Segur}(1981)}]{AblowitzSegur1981}%
  \BibitemOpen
  \bibfield  {author} {\bibinfo {author} {\bibfnamefont {M.~J.}\ \bibnamefont
  {Ablowitz}}\ and\ \bibinfo {author} {\bibfnamefont {H.}~\bibnamefont
  {Segur}},\ }\href@noop {} {\emph {\bibinfo {title} {Solitons and the inverse
  scattering transform}}}\ (\bibinfo  {publisher} {SIAM},\ \bibinfo {year}
  {1981})\BibitemShut {NoStop}%
\bibitem [{\citenamefont {Kodama}(2018)}]{Kodama2018}%
  \BibitemOpen
  \bibfield  {author} {\bibinfo {author} {\bibfnamefont {Y.}~\bibnamefont
  {Kodama}},\ }\href@noop {} {\emph {\bibinfo {title} {Solitons in
  two-dimensional shallow water}}}\ (\bibinfo  {publisher} {SIAM},\ \bibinfo
  {year} {2018})\BibitemShut {NoStop}%
\bibitem [{\citenamefont {Ablowitz}\ and\ \citenamefont
  {Segur}(1979)}]{ablowitz_evolution_1979}%
  \BibitemOpen
  \bibfield  {author} {\bibinfo {author} {\bibfnamefont {M.~J.}\ \bibnamefont
  {Ablowitz}}\ and\ \bibinfo {author} {\bibfnamefont {H.}~\bibnamefont
  {Segur}},\ }\href@noop {} {\bibfield  {journal} {\bibinfo  {journal} {J.
  Fluid Mech.}\ }\textbf {\bibinfo {volume} {92}},\ \bibinfo {pages} {691}
  (\bibinfo {year} {1979})}\BibitemShut {NoStop}%
\bibitem [{\citenamefont {Infeld}\ and\ \citenamefont
  {Rowlands}(2000)}]{InfeldRowlands}%
  \BibitemOpen
  \bibfield  {author} {\bibinfo {author} {\bibfnamefont {E.}~\bibnamefont
  {Infeld}}\ and\ \bibinfo {author} {\bibfnamefont {G.}~\bibnamefont
  {Rowlands}},\ }\href@noop {} {\emph {\bibinfo {title} {Nonlinear waves,
  solitons and chaos}}}\ (\bibinfo  {publisher} {Cambridge University Press},\
  \bibinfo {year} {2000})\BibitemShut {NoStop}%
\bibitem [{\citenamefont {Fal'kovich}\ and\ \citenamefont
  {Turitsyn}(1985)}]{JETP62p146}%
  \BibitemOpen
  \bibfield  {author} {\bibinfo {author} {\bibfnamefont {E.~G.}\ \bibnamefont
  {Fal'kovich}}\ and\ \bibinfo {author} {\bibfnamefont {S.~K.}\ \bibnamefont
  {Turitsyn}},\ }\href@noop {} {\bibfield  {journal} {\bibinfo  {journal} {Sov.
  Phys. JETP}\ }\textbf {\bibinfo {volume} {62}},\ \bibinfo {pages} {146}
  (\bibinfo {year} {1985})}\BibitemShut {NoStop}%
\bibitem [{\citenamefont {Leblond}(2002)}]{JPA35p10149}%
  \BibitemOpen
  \bibfield  {author} {\bibinfo {author} {\bibfnamefont {H.}~\bibnamefont
  {Leblond}},\ }\href@noop {} {\bibfield  {journal} {\bibinfo  {journal} {J.
  Phys. A}\ }\textbf {\bibinfo {volume} {35}},\ \bibinfo {pages} {10149}
  (\bibinfo {year} {2002})}\BibitemShut {NoStop}%
\bibitem [{\citenamefont {Kates}(1988)}]{AA194p3}%
  \BibitemOpen
  \bibfield  {author} {\bibinfo {author} {\bibfnamefont {R.}~\bibnamefont
  {Kates}},\ }\href@noop {} {\bibfield  {journal} {\bibinfo  {journal} {Astron.
  Astrophys.}\ }\textbf {\bibinfo {volume} {194}},\ \bibinfo {pages} {3}
  (\bibinfo {year} {1988})}\BibitemShut {NoStop}%
\bibitem [{\citenamefont {Vacaru}\ and\ \citenamefont
  {Singleton}(2002)}]{CQG19p2793}%
  \BibitemOpen
  \bibfield  {author} {\bibinfo {author} {\bibfnamefont {S.~I.}\ \bibnamefont
  {Vacaru}}\ and\ \bibinfo {author} {\bibfnamefont {D.}~\bibnamefont
  {Singleton}},\ }\href@noop {} {\bibfield  {journal} {\bibinfo  {journal}
  {Class. Quantum Grav.}\ }\textbf {\bibinfo {volume} {19}},\ \bibinfo {pages}
  {2793} (\bibinfo {year} {2002})}\BibitemShut {NoStop}%
\bibitem [{\citenamefont {Melnikov}\ \emph {et~al.}(2000)\citenamefont
  {Melnikov}, \citenamefont {Mihalache},\ and\ \citenamefont
  {Panoiu}}]{OC181p345}%
  \BibitemOpen
  \bibfield  {author} {\bibinfo {author} {\bibfnamefont {I.}~\bibnamefont
  {Melnikov}}, \bibinfo {author} {\bibfnamefont {D.}~\bibnamefont {Mihalache}},
  \ and\ \bibinfo {author} {\bibfnamefont {N.}~\bibnamefont {Panoiu}},\
  }\href@noop {} {\bibfield  {journal} {\bibinfo  {journal} {Opt. Commun.}\
  }\textbf {\bibinfo {volume} {181}},\ \bibinfo {pages} {345} (\bibinfo {year}
  {2000})}\BibitemShut {NoStop}%
\bibitem [{\citenamefont {Duncan}\ \emph {et~al.}(1991)\citenamefont {Duncan},
  \citenamefont {Eilbeck}, \citenamefont {Walshaw},\ and\ \citenamefont
  {Zakharov}}]{PLA158p107}%
  \BibitemOpen
  \bibfield  {author} {\bibinfo {author} {\bibfnamefont {D.}~\bibnamefont
  {Duncan}}, \bibinfo {author} {\bibfnamefont {J.}~\bibnamefont {Eilbeck}},
  \bibinfo {author} {\bibfnamefont {C.}~\bibnamefont {Walshaw}}, \ and\
  \bibinfo {author} {\bibfnamefont {V.}~\bibnamefont {Zakharov}},\ }\href@noop
  {} {\bibfield  {journal} {\bibinfo  {journal} {Phys. Lett. A}\ }\textbf
  {\bibinfo {volume} {158}},\ \bibinfo {pages} {107} (\bibinfo {year}
  {1991})}\BibitemShut {NoStop}%
\bibitem [{\citenamefont {Guo-Xiang}\ and\ \citenamefont
  {Shan-Hua}(2002)}]{CPL19p17}%
  \BibitemOpen
  \bibfield  {author} {\bibinfo {author} {\bibfnamefont {H.}~\bibnamefont
  {Guo-Xiang}}\ and\ \bibinfo {author} {\bibfnamefont {Z.}~\bibnamefont
  {Shan-Hua}},\ }\href@noop {} {\bibfield  {journal} {\bibinfo  {journal}
  {Chinese Phys. Lett.}\ }\textbf {\bibinfo {volume} {19}},\ \bibinfo {pages}
  {17} (\bibinfo {year} {2002})}\BibitemShut {NoStop}%
\bibitem [{\citenamefont {Berloff}(2002)}]{PRE65p174518}%
  \BibitemOpen
  \bibfield  {author} {\bibinfo {author} {\bibfnamefont {N.~G.}\ \bibnamefont
  {Berloff}},\ }\href@noop {} {\bibfield  {journal} {\bibinfo  {journal} {Phys.
  Rev. B}\ }\textbf {\bibinfo {volume} {65}},\ \bibinfo {pages} {174518}
  (\bibinfo {year} {2002})}\BibitemShut {NoStop}%
\bibitem [{\citenamefont {Berloff}(2004)}]{PRA69p53601}%
  \BibitemOpen
  \bibfield  {author} {\bibinfo {author} {\bibfnamefont {N.~G.}\ \bibnamefont
  {Berloff}},\ }\href@noop {} {\bibfield  {journal} {\bibinfo  {journal} {Phys.
  Rev. A}\ }\textbf {\bibinfo {volume} {69}},\ \bibinfo {pages} {053601}
  (\bibinfo {year} {2004})}\BibitemShut {NoStop}%
\bibitem [{\citenamefont {Novikov}\ \emph {et~al.}(1984)\citenamefont
  {Novikov}, \citenamefont {Manakov}, \citenamefont {Pitaevskii},\ and\
  \citenamefont {Zakharov}}]{NMPZ1984}%
  \BibitemOpen
  \bibfield  {author} {\bibinfo {author} {\bibfnamefont {S.~P.}\ \bibnamefont
  {Novikov}}, \bibinfo {author} {\bibfnamefont {S.~V.}\ \bibnamefont
  {Manakov}}, \bibinfo {author} {\bibfnamefont {L.~P.}\ \bibnamefont
  {Pitaevskii}}, \ and\ \bibinfo {author} {\bibfnamefont {V.~E.}\ \bibnamefont
  {Zakharov}},\ }\href@noop {} {\emph {\bibinfo {title} {Theory of solitons:
  the inverse scattering transform}}}\ (\bibinfo  {publisher} {Plenum},\
  \bibinfo {year} {1984})\BibitemShut {NoStop}%
\bibitem [{\citenamefont {Ablowitz}\ and\ \citenamefont
  {Clarkson}(1991)}]{AblowitzClarkson1991}%
  \BibitemOpen
  \bibfield  {author} {\bibinfo {author} {\bibfnamefont {M.~J.}\ \bibnamefont
  {Ablowitz}}\ and\ \bibinfo {author} {\bibfnamefont {P.~A.}\ \bibnamefont
  {Clarkson}},\ }\href@noop {} {\emph {\bibinfo {title} {Solitons, nonlinear
  evolution equations and inverse scattering}}}\ (\bibinfo  {publisher}
  {Cambridge},\ \bibinfo {year} {1991})\BibitemShut {NoStop}%
\bibitem [{\citenamefont {Konopelchenko}(1993)}]{Konopelchenko1993}%
  \BibitemOpen
  \bibfield  {author} {\bibinfo {author} {\bibfnamefont {B.~G.}\ \bibnamefont
  {Konopelchenko}},\ }\href@noop {} {\emph {\bibinfo {title} {Solitons in
  multidimensions: inverse spectral transform method}}}\ (\bibinfo  {publisher}
  {World Scientific},\ \bibinfo {year} {1993})\BibitemShut {NoStop}%
\bibitem [{\citenamefont {Biondini}\ and\ \citenamefont
  {Kodama}(2003)}]{JPA36p10519}%
  \BibitemOpen
  \bibfield  {author} {\bibinfo {author} {\bibfnamefont {G.}~\bibnamefont
  {Biondini}}\ and\ \bibinfo {author} {\bibfnamefont {Y.}~\bibnamefont
  {Kodama}},\ }\href@noop {} {\bibfield  {journal} {\bibinfo  {journal} {J.
  Phys. A}\ }\textbf {\bibinfo {volume} {36}},\ \bibinfo {pages} {10519}
  (\bibinfo {year} {2003})}\BibitemShut {NoStop}%
\bibitem [{\citenamefont {Biondini}\ and\ \citenamefont
  {Chakravarty}(2006)}]{JMP47p033514}%
  \BibitemOpen
  \bibfield  {author} {\bibinfo {author} {\bibfnamefont {G.}~\bibnamefont
  {Biondini}}\ and\ \bibinfo {author} {\bibfnamefont {S.}~\bibnamefont
  {Chakravarty}},\ }\href@noop {} {\bibfield  {journal} {\bibinfo  {journal}
  {J. Math. Phys.}\ }\textbf {\bibinfo {volume} {47}},\ \bibinfo {pages}
  {033514} (\bibinfo {year} {2006})}\BibitemShut {NoStop}%
\bibitem [{\citenamefont {Biondini}(2007)}]{PRL99p064103}%
  \BibitemOpen
  \bibfield  {author} {\bibinfo {author} {\bibfnamefont {G.}~\bibnamefont
  {Biondini}},\ }\href@noop {} {\bibfield  {journal} {\bibinfo  {journal}
  {Phys. Rev. Lett.}\ }\textbf {\bibinfo {volume} {99}},\ \bibinfo {pages}
  {064103} (\bibinfo {year} {2007})}\BibitemShut {NoStop}%
\bibitem [{\citenamefont {Kodama}(2017)}]{Kodama2017}%
  \BibitemOpen
  \bibfield  {author} {\bibinfo {author} {\bibfnamefont {Y.}~\bibnamefont
  {Kodama}},\ }\href@noop {} {\emph {\bibinfo {title} {{KP} solitons and the
  {Grassmannians}: {Combinatorics} and geometry of two-dimensional wave
  patterns}}}\ (\bibinfo  {publisher} {Springer},\ \bibinfo {year}
  {2017})\BibitemShut {NoStop}%
\bibitem [{\citenamefont {Whitham}(1974)}]{Whitham1974}%
  \BibitemOpen
  \bibfield  {author} {\bibinfo {author} {\bibfnamefont {G.~B.}\ \bibnamefont
  {Whitham}},\ }\href@noop {} {\emph {\bibinfo {title} {Linear and nonlinear
  waves}}}\ (\bibinfo  {publisher} {Wiley},\ \bibinfo {year}
  {1974})\BibitemShut {NoStop}%
\bibitem [{\citenamefont {Ryskamp}\ \emph
  {et~al.}(2021{\natexlab{a}})\citenamefont {Ryskamp}, \citenamefont {Hoefer},\
  and\ \citenamefont {Biondini}}]{NLTY2021v34p3583}%
  \BibitemOpen
  \bibfield  {author} {\bibinfo {author} {\bibfnamefont {S.}~\bibnamefont
  {Ryskamp}}, \bibinfo {author} {\bibfnamefont {M.~A.}\ \bibnamefont {Hoefer}},
  \ and\ \bibinfo {author} {\bibfnamefont {G.}~\bibnamefont {Biondini}},\
  }\href@noop {} {\bibfield  {journal} {\bibinfo  {journal} {Nonlinearity}\
  }\textbf {\bibinfo {volume} {34}},\ \bibinfo {pages} {3583} (\bibinfo {year}
  {2021}{\natexlab{a}})}\BibitemShut {NoStop}%
\bibitem [{\citenamefont {Olver}\ \emph {et~al.}(2010)\citenamefont {Olver},
  \citenamefont {Lozier}, \citenamefont {Boisvert},\ and\ \citenamefont
  {Clark}}]{NIST2010}%
  \BibitemOpen
  \bibfield  {author} {\bibinfo {author} {\bibfnamefont {F.~W.}\ \bibnamefont
  {Olver}}, \bibinfo {author} {\bibfnamefont {D.~W.}\ \bibnamefont {Lozier}},
  \bibinfo {author} {\bibfnamefont {R.~F.}\ \bibnamefont {Boisvert}}, \ and\
  \bibinfo {author} {\bibfnamefont {C.~W.}\ \bibnamefont {Clark}},\ }\href@noop
  {} {\emph {\bibinfo {title} {NIST Handbook of Mathematical Functions}}}\
  (\bibinfo  {publisher} {Cambridge University Press},\ \bibinfo {year}
  {2010})\BibitemShut {NoStop}%
\bibitem [{\citenamefont {Ablowitz}\ \emph {et~al.}(2017)\citenamefont
  {Ablowitz}, \citenamefont {Biondini},\ and\ \citenamefont
  {Wang}}]{RSPA2017v473p20160695}%
  \BibitemOpen
  \bibfield  {author} {\bibinfo {author} {\bibfnamefont {M.~J.}\ \bibnamefont
  {Ablowitz}}, \bibinfo {author} {\bibfnamefont {G.}~\bibnamefont {Biondini}},
  \ and\ \bibinfo {author} {\bibfnamefont {Q.}~\bibnamefont {Wang}},\
  }\href@noop {} {\bibfield  {journal} {\bibinfo  {journal} {Roy. Soc. Proc.
  A}\ }\textbf {\bibinfo {volume} {473}},\ \bibinfo {pages} {20160695}
  (\bibinfo {year} {2017})}\BibitemShut {NoStop}%
\bibitem [{sup()}]{supplement}%
  \BibitemOpen
  \href@noop {} {}\bibinfo {howpublished} {See the supplementary material for
  further details.}\BibitemShut {Stop}%
\bibitem [{\citenamefont {Gurevich}\ and\ \citenamefont
  {Pitaevskii}(1974)}]{JETP38p291}%
  \BibitemOpen
  \bibfield  {author} {\bibinfo {author} {\bibfnamefont {A.~V.}\ \bibnamefont
  {Gurevich}}\ and\ \bibinfo {author} {\bibfnamefont {L.~P.}\ \bibnamefont
  {Pitaevskii}},\ }\href@noop {} {\bibfield  {journal} {\bibinfo  {journal}
  {Zh. Eksp. Teor. Fiz.}\ }\textbf {\bibinfo {volume} {65}},\ \bibinfo {pages}
  {590} (\bibinfo {year} {1974})}\BibitemShut {NoStop}%
\bibitem [{\citenamefont {Segur}\ and\ \citenamefont
  {Finkel}(1985)}]{SAPM73p183}%
  \BibitemOpen
  \bibfield  {author} {\bibinfo {author} {\bibfnamefont {H.}~\bibnamefont
  {Segur}}\ and\ \bibinfo {author} {\bibfnamefont {A.}~\bibnamefont {Finkel}},\
  }\href@noop {} {\bibfield  {journal} {\bibinfo  {journal} {Stud. Appl.
  Math.}\ }\textbf {\bibinfo {volume} {73}},\ \bibinfo {pages} {183} (\bibinfo
  {year} {1985})}\BibitemShut {NoStop}%
\bibitem [{\citenamefont {Baines}(1998)}]{baines_topographic_1998}%
  \BibitemOpen
  \bibfield  {author} {\bibinfo {author} {\bibfnamefont {P.~G.}\ \bibnamefont
  {Baines}},\ }\href@noop {} {\emph {\bibinfo {title} {Topographic {{Effects}}
  in {{Stratified Flows}}}}}\ (\bibinfo  {publisher} {Cambridge University
  Press},\ \bibinfo {address} {Cambridge, UK},\ \bibinfo {year}
  {1998})\BibitemShut {NoStop}%
\bibitem [{\citenamefont {Ryskamp}\ \emph
  {et~al.}(2021{\natexlab{b}})\citenamefont {Ryskamp}, \citenamefont {Maiden},
  \citenamefont {Hoefer},\ and\ \citenamefont {Biondini}}]{JFM2021v909pA24}%
  \BibitemOpen
  \bibfield  {author} {\bibinfo {author} {\bibfnamefont {S.}~\bibnamefont
  {Ryskamp}}, \bibinfo {author} {\bibfnamefont {M.}~\bibnamefont {Maiden}},
  \bibinfo {author} {\bibfnamefont {M.~A.}\ \bibnamefont {Hoefer}}, \ and\
  \bibinfo {author} {\bibfnamefont {G.}~\bibnamefont {Biondini}},\ }\href@noop
  {} {\bibfield  {journal} {\bibinfo  {journal} {J.\ Fluid Mech.}\ }\textbf
  {\bibinfo {volume} {909}},\ \bibinfo {pages} {A24} (\bibinfo {year}
  {2021}{\natexlab{b}})}\BibitemShut {NoStop}%
\bibitem [{\citenamefont {Ryskamp}\ \emph {et~al.}(2022)\citenamefont
  {Ryskamp}, \citenamefont {Hoefer},\ and\ \citenamefont
  {Biondini}}]{RSPA2022v478v20210823}%
  \BibitemOpen
  \bibfield  {author} {\bibinfo {author} {\bibfnamefont {S.}~\bibnamefont
  {Ryskamp}}, \bibinfo {author} {\bibfnamefont {M.~A.}\ \bibnamefont {Hoefer}},
  \ and\ \bibinfo {author} {\bibfnamefont {G.}~\bibnamefont {Biondini}},\
  }\href@noop {} {\bibfield  {journal} {\bibinfo  {journal} {Roy.\ Soc.\ Proc.\
  A}\ }\textbf {\bibinfo {volume} {478}},\ \bibinfo {pages} {20210823}
  (\bibinfo {year} {2022})}\BibitemShut {NoStop}%
\bibitem [{\citenamefont {Li}\ \emph {et~al.}(2011)\citenamefont {Li},
  \citenamefont {Yeh},\ and\ \citenamefont {Kodama}}]{JFM672p326}%
  \BibitemOpen
  \bibfield  {author} {\bibinfo {author} {\bibfnamefont {W.}~\bibnamefont
  {Li}}, \bibinfo {author} {\bibfnamefont {H.}~\bibnamefont {Yeh}}, \ and\
  \bibinfo {author} {\bibfnamefont {Y.}~\bibnamefont {Kodama}},\ }\href@noop {}
  {\bibfield  {journal} {\bibinfo  {journal} {Journal of fluid mechanics}\
  }\textbf {\bibinfo {volume} {672}},\ \bibinfo {pages} {326} (\bibinfo {year}
  {2011})}\BibitemShut {NoStop}%
\bibitem [{\citenamefont {Mach}\ and\ \citenamefont
  {Wosyka}(1875)}]{mach_1875}%
  \BibitemOpen
  \bibfield  {author} {\bibinfo {author} {\bibfnamefont {E.}~\bibnamefont
  {Mach}}\ and\ \bibinfo {author} {\bibfnamefont {J.}~\bibnamefont {Wosyka}},\
  }\href@noop {} {\bibfield  {journal} {\bibinfo  {journal} {Sitzungsber. Akad.
  Wiss. Wien}\ }\textbf {\bibinfo {volume} {72}},\ \bibinfo {pages} {44}
  (\bibinfo {year} {1875})}\BibitemShut {NoStop}%
\bibitem [{\citenamefont {Krehl}\ and\ \citenamefont {{van der
  Geest}}(1991)}]{krehl_discovery_1991}%
  \BibitemOpen
  \bibfield  {author} {\bibinfo {author} {\bibfnamefont {P.}~\bibnamefont
  {Krehl}}\ and\ \bibinfo {author} {\bibfnamefont {M.}~\bibnamefont {{van der
  Geest}}},\ }\href@noop {} {\bibfield  {journal} {\bibinfo  {journal} {Shock
  Waves}\ }\textbf {\bibinfo {volume} {1}},\ \bibinfo {pages} {3} (\bibinfo
  {year} {1991})}\BibitemShut {NoStop}%
\bibitem [{\citenamefont {Lee}\ and\ \citenamefont
  {Grimshaw}(1990)}]{lee_upstreamadvancing_1990}%
  \BibitemOpen
  \bibfield  {author} {\bibinfo {author} {\bibfnamefont {S.-J.}\ \bibnamefont
  {Lee}}\ and\ \bibinfo {author} {\bibfnamefont {R.~H.~J.}\ \bibnamefont
  {Grimshaw}},\ }\href@noop {} {\bibfield  {journal} {\bibinfo  {journal}
  {Physics of Fluids A: Fluid Dynamics}\ }\textbf {\bibinfo {volume} {2}},\
  \bibinfo {pages} {194} (\bibinfo {year} {1990})}\BibitemShut {NoStop}%
\bibitem [{\citenamefont {Li}\ and\ \citenamefont
  {Sclavounos}(2002)}]{li_three-dimensional_2002}%
  \BibitemOpen
  \bibfield  {author} {\bibinfo {author} {\bibfnamefont {Y.}~\bibnamefont
  {Li}}\ and\ \bibinfo {author} {\bibfnamefont {P.~D.}\ \bibnamefont
  {Sclavounos}},\ }\href@noop {} {\bibfield  {journal} {\bibinfo  {journal} {J.
  Fluid Mech.}\ }\textbf {\bibinfo {volume} {470}},\ \bibinfo {pages} {383}
  (\bibinfo {year} {2002})}\BibitemShut {NoStop}%
\bibitem [{\citenamefont {Soomere}(2007)}]{soomere_nonlinear_2007}%
  \BibitemOpen
  \bibfield  {author} {\bibinfo {author} {\bibfnamefont {T.}~\bibnamefont
  {Soomere}},\ }\href@noop {} {\bibfield  {journal} {\bibinfo  {journal}
  {Applied Mechanics Reviews}\ }\textbf {\bibinfo {volume} {60}},\ \bibinfo
  {pages} {120} (\bibinfo {year} {2007})}\BibitemShut {NoStop}%
\bibitem [{\citenamefont {Mann}(2019)}]{mann_improve_2019}%
  \BibitemOpen
  \bibfield  {author} {\bibinfo {author} {\bibfnamefont {A.}~\bibnamefont
  {Mann}},\ }\href@noop {} {\bibfield  {journal} {\bibinfo  {journal} {Proc.
  Natl. Acad. Sci.}\ }\textbf {\bibinfo {volume} {116}},\ \bibinfo {pages}
  {19218} (\bibinfo {year} {2019})}\BibitemShut {NoStop}%
\bibitem [{\citenamefont {Yuan}\ and\ \citenamefont
  {Wang}(2022)}]{yuan_diffraction_2022}%
  \BibitemOpen
  \bibfield  {author} {\bibinfo {author} {\bibfnamefont {C.}~\bibnamefont
  {Yuan}}\ and\ \bibinfo {author} {\bibfnamefont {Z.}~\bibnamefont {Wang}},\
  }\href@noop {} {\bibfield  {journal} {\bibinfo  {journal} {J. Fluid Mech.}\
  }\textbf {\bibinfo {volume} {936}},\ \bibinfo {pages} {A20} (\bibinfo {year}
  {2022})}\BibitemShut {NoStop}%
\bibitem [{\citenamefont {Grava}\ and\ \citenamefont
  {Tian}(2002)}]{CPAM55p1569}%
  \BibitemOpen
  \bibfield  {author} {\bibinfo {author} {\bibfnamefont {T.}~\bibnamefont
  {Grava}}\ and\ \bibinfo {author} {\bibfnamefont {F.}~\bibnamefont {Tian}},\
  }\href@noop {} {\bibfield  {journal} {\bibinfo  {journal} {Commun. Pure Appl.
  Math.}\ }\textbf {\bibinfo {volume} {55}},\ \bibinfo {pages} {1569} (\bibinfo
  {year} {2002})}\BibitemShut {NoStop}%
\bibitem [{\citenamefont {Bonnemain}\ \emph {et~al.}(2024)\citenamefont
  {Bonnemain}, \citenamefont {Biondini}, \citenamefont {Doyon}, \citenamefont
  {Roberti},\ and\ \citenamefont {El}}]{arxiv2408.05548}%
  \BibitemOpen
  \bibfield  {author} {\bibinfo {author} {\bibfnamefont {T.}~\bibnamefont
  {Bonnemain}}, \bibinfo {author} {\bibfnamefont {G.}~\bibnamefont {Biondini}},
  \bibinfo {author} {\bibfnamefont {B.}~\bibnamefont {Doyon}}, \bibinfo
  {author} {\bibfnamefont {G.}~\bibnamefont {Roberti}}, \ and\ \bibinfo
  {author} {\bibfnamefont {G.~A.}\ \bibnamefont {El}},\ }\href@noop {}
  {\enquote {\bibinfo {title} {Two-dimensional stationary soliton gas},}\ }
  (\bibinfo {year} {2024})\BibitemShut {NoStop}%
\bibitem [{\citenamefont {Zheng}(2001)}]{Zheng2001}%
  \BibitemOpen
  \bibfield  {author} {\bibinfo {author} {\bibfnamefont {Y.}~\bibnamefont
  {Zheng}},\ }\href@noop {} {\emph {\bibinfo {title} {Systems of conservation
  laws: Two-dimensional Riemann problems}}}\ (\bibinfo  {publisher}
  {Springer},\ \bibinfo {year} {2001})\BibitemShut {NoStop}%
\bibitem [{\citenamefont {Krichever}(1988)}]{Krichever}%
  \BibitemOpen
  \bibfield  {author} {\bibinfo {author} {\bibfnamefont {I.~M.}\ \bibnamefont
  {Krichever}},\ }\href@noop {} {\bibfield  {journal} {\bibinfo  {journal}
  {Funct. Anal. Appl.}\ }\textbf {\bibinfo {volume} {22}},\ \bibinfo {pages}
  {37} (\bibinfo {year} {1988})}\BibitemShut {NoStop}%
\bibitem [{\citenamefont {Hoefer}\ and\ \citenamefont
  {Ablowitz}(2007)}]{PHYSD236p44}%
  \BibitemOpen
  \bibfield  {author} {\bibinfo {author} {\bibfnamefont {M.~A.}\ \bibnamefont
  {Hoefer}}\ and\ \bibinfo {author} {\bibfnamefont {M.~J.}\ \bibnamefont
  {Ablowitz}},\ }\href@noop {} {\bibfield  {journal} {\bibinfo  {journal}
  {Phys. D}\ }\textbf {\bibinfo {volume} {236}},\ \bibinfo {pages} {44}
  (\bibinfo {year} {2007})}\BibitemShut {NoStop}%
\bibitem [{\citenamefont {Klein}\ \emph {et~al.}(2007)\citenamefont {Klein},
  \citenamefont {Sparber},\ and\ \citenamefont {Markowich}}]{JNLSCI17p429}%
  \BibitemOpen
  \bibfield  {author} {\bibinfo {author} {\bibfnamefont {C.}~\bibnamefont
  {Klein}}, \bibinfo {author} {\bibfnamefont {C.}~\bibnamefont {Sparber}}, \
  and\ \bibinfo {author} {\bibfnamefont {P.}~\bibnamefont {Markowich}},\
  }\href@noop {} {\bibfield  {journal} {\bibinfo  {journal} {J. Nonl. Sci.}\
  }\textbf {\bibinfo {volume} {17}},\ \bibinfo {pages} {429} (\bibinfo {year}
  {2007})}\BibitemShut {NoStop}%
\bibitem [{\citenamefont {Liu}\ and\ \citenamefont
  {Trogdon}(2023)}]{LiuTrogdon}%
  \BibitemOpen
  \bibfield  {author} {\bibinfo {author} {\bibfnamefont {A.}~\bibnamefont
  {Liu}}\ and\ \bibinfo {author} {\bibfnamefont {T.}~\bibnamefont {Trogdon}},\
  }\href@noop {} {\bibfield  {journal} {\bibinfo  {journal} {Appl. Numer.
  Math.}\ }\textbf {\bibinfo {volume} {192}},\ \bibinfo {pages} {19} (\bibinfo
  {year} {2023})}\BibitemShut {NoStop}%
\end{thebibliography}%
